\documentclass[12pt,preprint]{aastex}

\newcommand{\Lya}{\hbox{Ly$\alpha$}}

\newcommand{\Jup}{\hbox{$J^\prime$}}
\newcommand{\Jlo}{\hbox{$J^{\prime\prime}$}}
\newcommand{\Vup}{\hbox{$v^\prime$}}
\newcommand{\Vlo}{\hbox{$v^{\prime\prime}$}}
\newcommand{\Elo}{\hbox{$E^{\prime\prime}$}}
\newcommand{\kms}{\hbox{km s$^{-1}$}}
\newcommand{\erg}{\hbox{erg cm$^{-2}$ s$^{-1}$}}

\newcommand{\mdotyr}{\hbox{$M_\odot$ yr$^{-1}$}}
\newcommand{\IUE}{\textit{IUE}}

\newcommand{\HST}{\textit{HST}}
\newcommand{\STIS}{STIS}            
\newcommand{\FUSE}{\textit{FUSE}}

\slugcomment{accepted by ApJS}
\shorttitle{Origins of Fluorescent H$_2$ Emission}
\shortauthors{Herczeg et al.}

\begin{document}

\title{The Origins of Fluorescent H$_2$ Emission From T~Tauri~Stars}


\author{Gregory J. Herczeg\altaffilmark{1,2}, Jeffrey L. Linsky\altaffilmark{1}, 
Frederick M. Walter\altaffilmark{3}, G\"osta F. Gahm\altaffilmark{4}, Christopher M. Johns-Krull\altaffilmark{5}}

\altaffiltext{1}{JILA,  University of Colorado and NIST, Boulder, CO 80309-0440;  gregoryh@astro.caltech.edu, jlinsky@jila.colorado.edu}

\altaffiltext{2}{Current address Caltech, MC105-24, 1200 E. California Blvd., Pasadena, CA 91125}

\altaffiltext{3}{Department of Physics and Astronomy, Stony Brook University, Stony
Brook NY 11794-3800; fwalter@astro.sunysb.edu, msimon@astro.sunysb.edu}

\altaffiltext{4}{Stockholm Observatory, AlbaNova, SE - 106 91 Stockholm, Sweden; gahm@astro.su.se}

\altaffiltext{5}{Department of Physics \& Astronomy, Rice University, Houston TX
77005-1892; cmj@rice.edu}

\begin{abstract}
We survey fluorescent H$_2$ emission in \HST/STIS spectra of the 
classical T Tauri stars (CTTSs) TW Hya, DF Tau, RU Lupi, T Tau, 
and DG Tau, and the weak-lined T Tauri star (WTTS) V836 Tau.
From each of those sources we detect between
41--209 narrow H$_2$ emission lines, most of which are pumped by strong \Lya\ emission.
H$_2$ emission is not detected from the WTTS V410 Tau.  
The fluorescent H$_2$ emission appears to be common to circumstellar 
environments around all CTTSs, but high spectral and spatial resolution 
STIS observations reveal diverse phenomenon.  
Blueshifted H$_2$ emission detected from RU Lupi, T Tau, 
and DG Tau is consistent with an origin in an outflow.
The H$_2$ emission from TW Hya, DF Tau, 
and V836 Tau is centered at the radial velocity of the star and is consistent 
with an origin in a warm disk surface.
The H$_2$ lines from RU Lupi, DF
Tau, and T Tau also have excess blueshifted H$_2$ emission that extends to
as much as -100 \kms.  The strength of this blueshifted component from DF
Tau and T Tau depends on the upper level of the transition. 
In all cases, the small aperture and attenuation of H$_2$
emission by stellar winds restricts the H$_2$ 
emission to be formed close to the star.  The \Lya\ and the H$_2$ emission
blueshifted by 15 \kms\ relative to RU Lupi are extended to the SW by $\sim 0\farcs07$, although the faster H$_2$ gas that extends to $\sim 100$ \kms\ is
not spatially extended.  We also find a
small reservoir of H$_2$ emission from TW Hya and DF Tau consistent with an 
excitation temperature of $\sim2.5\times10^4$ K.
\end{abstract}


%
%


\section{INTRODUCTION}
Molecular hydrogen is prevalent in both circumstellar disks and nebulosity
around young stars.  Observationally discriminating between these two
sources of H$_2$ gas could provide a valuable probe of the physical
characteristics and evolution of gas in protoplanetary disks.  While other
probes of this gas, such as CO and H$_2$O, have yielded powerful insights into the
physical conditions of the disk \citep[e.g.,][]{Naj03,Bri03,Car04},
identifying H$_2$ emission from the disk has been difficult
because IR rovibrational transitions are weak, cold H$_2$ does not radiate,
and diagnostics of H$_2$ gas in the disk can be contaminated by H$_2$
in surrounding molecular gas.

A variety of methods involving H$_2$ emission have been used to probe the circumstellar 
environments around young stars.  H$_2$ emission was first detected around 
a young star in IR observations of the 1-0 S(1) line at 2.1218 $\mu$m from T Tau by \citet{Bec78}.  
\citet{Bro81} used \IUE\ to
detect far-ultraviolet (FUV, $\lambda<2000$ \AA) H$_2$ emission from T Tau.  
IR maps of emission in the 1-0 S(1) line and long-slit FUV spectra of H$_2$
fluorescence reveal that the
hot gas extends to 20$^{\prime\prime}$ from T Tau and is most likely heated by stellar outflows 
that shock molecular material near the stars \citep{Van94,Wal03,Sau03}.

\citet{Val00} detected Lyman-band H$_2$ emission in 13 of 32 classical T Tauri stars (CTTSs) observed
in low-resolution ($R\equiv\frac{\lambda}{\Delta \lambda}\sim200$) FUV
spectra obtained with {\it IUE} and
suggested that most of the non-detections resulted from inadequate sensitivity.
\citet{Ard02a} found in \HST/GHRS spectra ($R=20,000$) of eight CTTSs that 
the H$_2$ lines are blueshifted by 0--20 \kms\ relative to
the radial velocity of the star.
They note that systematic uncertainties in the wavelength calibration of GHRS could 
be as large as 20 \kms\ for several of their observations that occurred before
COSTAR was installed on {\it HST}.  However, they  use the absence
of any stars with redshifted H$_2$ emission to suggest 
that, for some sources, the blueshift may be significant.  This H$_2$
emission would therefore be produced by stellar outflows. 
The limited spectral coverage and large $(\sim 2^{\prime\prime})$ aperture 
used in the GHRS observations
prevented a thorough analysis of the H$_2$ lines, but \citet{Ard02a}
confirmed that these lines are
pumped by \Lya.  Fluorescent H$_2$ emission is also found from the accreting
brown dwarf 2MASS J1207334-393254 \citep{Giz05}.
\citet{Bar03} detected warm H$_2$
in emission in the 1-0 S(1) line
at 2.1218 $\mu$m from three of five CTTSs and one of 11 weak-lined T Tauri star (WTTSs).  
Based on kinematics, they suggested that this H$_2$ emission is produced in the disk
within 30 AU of the central star.  

Emission in the 2.1218 $\mu$m line and the FUV lines is produced by
warm (1000--3000 K) gas but not by the cold gas that comprises the bulk of
the mass in circumstellar disks.
Although H$_2$ does not radiate at temperatures of $\sim 10$ K, emission from 50--100 K gas can 
be detected in pure rotational
H$_2$ lines.  \citet{Thi01} used {\it ISO} to 
detect emission in the pure rotational H$_2$
S(1) and S(0) lines at 17 and 28 $\mu$m, respectively.  
However, \citet{Ric02}, \citet{She03}, and \citet{Sak05} did not detect
emission in the S(2) 12 $\mu$m line or the S(1) 17 $\mu$m line from many
young stars in their ground-based observations that used much smaller apertures
than {\it ISO}.  Several of these sources had claimed {\it ISO} detections
of H$_2$, even though the ground-based non-detections were more sensitive
to H$_2$ in a disk than {\it ISO}.
  Midway into the analysis of a larger sample, \citet{Ric04} 
reported detections 
of these lines from several young stars including T Tau. 
If the {\it ISO} detections are real,
then this H$_2$ emission is produced 
in a molecular cloud 
or an envelope extended beyond the circumstellar disk.  

Cold H$_2$ gas can also be observed in absorption from the ground
vibrational level at $\lambda<1120$ \AA.  However, the
detection of H$_2$ absorption through a disk requires that the disk be viewed
nearly edge-on yet also be optically thin to FUV emission. 
The prevalence of H$_2$ absorption toward
Herbig AeBe (HAeBe) stars observed with {\it FUSE}  \citep{Rob01,Lec03,Bou03,Gra04,Mar04,Mar05}
suggests that the H$_2$ absorption toward these more massive stars occurs
in a molecular cloud or
remnant molecular envelope rather than in a disk.  For example, based on the 
observed radial velocity \citet{Mar05} suggest that the H$_2$
absorption from HD141569 is related to the nearby dark cloud L134N.
\citet{Lec01} and
\citet{Rob05} use the absence of any H$_2$ absorption in {\it FUSE} spectra
to place strong upper limits on the molecular gas mass in the evolved,
optically thin disks of  $\beta$ Pic and AU Mic, respectively, which are
both observed nearly edge-on.

The E140M echelle spectrograph on \HST/STIS provides high resolution spectra 
covering a large wavelength range, 
permiting a detailed analysis of fluorescent H$_2$ emission in the FUV.  
In a STIS spectrum of TW Hya, \citet{Her02} detected over 140 H$_2$ lines from 19
distinct upper levels, demonstrating that the \Lya\ emission that pumps this
H$_2$ emission is broad.  The characteristics of this emission and the lack
of any other molecular gas 
near TW Hya suggest that the emission was produced at the disk surface, in a
thin layer heated to $T\sim2500$ K \citep{Her02,Her04}.
Although current models of gas in the disk suggest that neither FUV nor
X-ray irradiation alone can produce temperatures of 2500 K
\citep{Gla04,Nom05}, together they may sufficiently heat the gas to
explain the FUV H$_2$ emission. 

In contrast, long-slit \HST/STIS spectra of T Tau reveal extended H$_2$
emission from only two upper levels that are pumped close to \Lya\ line
center \citep{Wal03,Sau03}, which is similar to the H$_2$ fluorescence
detected toward HH43 and HH47 \citep{Sch83,Cur95}.  The on-source spectrum
of T Tau shows a much richer
H$_2$ spectrum, similar to that observed toward TW Hya, than is observed
off-source \citep{Wal03}. 

The observed FUV spectra of 
CTTSs are dominated by strong emission in lines of \ion{C}{4}, 
\ion{Si}{4}, \ion{C}{2}, \ion{O}{1}, and 
many Ly$\alpha$-pumped H$_2$ lines.  Strong 
\Lya\ emission is observed if the small \ion{H}{1} column density in our line of sight to the star is small.
In addition to these lines, the FUV continuum 
emission from TW Hya rises at $\lambda<1650$ \AA\ and is significantly 
enhanced above the accretion continuum that dominates the NUV emission 
from CTTSs \citep{Her04,Ber04}.  \citet{Ber04} also detected this emission from DM Tau, GM Aur, 
and LkCa15.  They identify this continuum or pseudo-continuum 
emission as H$_2$,
probably produced by collisions with energetic electrons, as may be seen 
from HH1/2 \citep{Ray97}.  \citet{Ber04} speculate
that this FUV continuum may be related to a deficiency of disk emission at
$\lambda<10$ $\mu$m, which is caused by the absence of optically thick
micron-sized dust within a few AU of the central star \citep{Cal02,Dal05}.
The ongoing accretion requires that the gas in the disk near the star is
still present.
The source of this H$_2$ continuum emission could be the disk surface but
this remains to be explored.  \citet{Her05} found that the FUV continuum is
weak or not present from RU
Lupi, even though RU Lupi is a source of strong fluorescent H$_2$ emission.

In this paper, we survey and analyze H$_2$ emission in the FUV spectra of five CTTSs and one
weakly accreting WTTS.  We report non-detections of H$_2$ emission in the spectra of
two WTTSs, only one of which is marginally significant.  
In \S 2 we describe our observations and in \S 3 we describe our targets.
In \S 4 we present an overview of FUV H$_2$ emission from each source,
including line fluxes and pumping mechanisms.  In \S 5 and \S 6 we analyze the
properties of the H$_2$ emission, including spectral and spatial emission
profiles, and consider  the origin of this emission.  
\S 7 summarizes our conclusions.
The H$_2$ emission
from some sources is consistent with a disk origin, but from other sources
is consistent with an outflow origin.  High spectral and
spatial resolution is essential to understanding H$_2$ emission from young
stars.

\section{OBSERVATIONS}
We observed the T Tauri stars DF Tau, RU Lupi, T Tau, DG Tau, V836 Tau,
V410 Tau, and V819 Tau with \HST/STIS as part of \HST\ program GO-8157. 
Each FUV observation consists of 4--5 orbits using the E140M echelle
spectrograph, spanning 1170--1710 \AA, with the $0\farcs2\times0\farcs06$
aperture to
isolate on-source emission.  Each visit 
includes a brief long-slit optical spectrum with the G430L grating,
spanning 2850--5750 \AA.  This program also included several long-slit STIS
FUV spectra 
of T Tau, which were analyzed by \citet{Wal03}.  We 
obtained FUV, NUV, and optical STIS spectra of DF Tau as part of \HST\
program GTO-7718.  This FUV observation of DF Tau used the unsupported
$0\farcs5\times0\farcs5$ aperture, which reduces the spectral resolution
but provides more
spatial information.  We also include and further analyze the \HST/STIS FUV spectrum
of TW Hya \citep{Her02} that was  obtained
as program GTO-8041.  Details of these observations
are listed in Table 1.

The stars observed with the $0\farcs2\times0\farcs06$ aperture were acquired in a 1s exposure 
using the F28x50LP optical long pass filter that includes 5500--10000
\AA, with a peak sensitivity at 6000 \AA.  
 We then peaked up on the sources using an optical white-light mirror and
 the CCD with the $0\farcs2\times0\farcs06$ aperture.  
For our observations of TW Hya and DF Tau that used the
$0\farcs5\times0\farcs5$ aperture, the stars were acquired with a
narrowband filter ($\sim90$ \AA) centered on the [\ion{O}{2}] 3727 \AA\
line.  For both TW Hya and DF Tau, the emission in this bandpass consists of the accretion continuum, high Balmer lines also produced by accreting gas, and
weak photospheric emission.  No peakup was necessary for these
observations.

The pixel size of the FUV MAMA detectors in the E140M echelle mode is
0\farcs036 ($\sim 3.3$ \kms) in the dispersion direction and 0\farcs029
(1.7 AU at the 57 pc distance of TW Hya and 4 AU at the $\sim140$ pc
distance of the other stars in our sample) in the cross-dispersion
direction.

We reduced the spectra using the {\it calSTIS} pipeline written in {\it IDL}
\citep{Lin99}.  We corrected the FUV echelle spectra for scattered light
using the {\it echelle\_scat} routine in {\it IDL}.  Several steps described below required
manual processing.

The automated pipeline processing did not successfully
extract the weak spectrum from our observation of DG Tau.  We found
the spectrum on the detector by searching for the maximum flux in several H$_2$ lines
in the extraction window.  Our extraction window is large enough to include
any stellar emission, even if the H$_2$ emission is extended only to one 
side of DG Tau.

As the telescope breathes, the thermal focus changes can modulate the count
rate when the point-spread function is 
larger than the
aperture.   
The observations of RU Lupi and T Tau obtained with the
$0\farcs2\times0\farcs06$ aperture both exhibit increasing count rates 
during each orbit that correlate with the improving telescope focus.
We calibrate the flux of T Tau following the method \citet{Her05} developed
for RU Lup. They noted a similar flux 
          increase in E140M observations of the continuum FUV source V471 Tau
          ({\it HST} programs GO-7735 and GO-9283, P.I. F. Walter), which were 
observed through the same
          small aperture. The $0\farcs06$ slit width is comparable to the width of
          the PSF, and \citet{Her05} showed that the
          flux correlates well with the instantaneous FWHM of the target
          in the cross-dispersion direction.
We measured the flux and
FWHM of emission in the cross-dispersion 
direction for several
spectral regions in 300 s intervals.  
These spectral regions are dominated by 
strong lines of \ion{C}{2} and \ion{C}{4} that are 
produced by accreting gas and are not extended beyond a 
point source.  The spectral regions do not include a significant 
contribution from H$_2$ emission, which may be extended (see \S 5.2).  
Figure 1 compares the flux and FWHM in four regions of the T Tau spectrum 
with those from V471 Tau. 
We calculate a flux from T Tau of
$1.82\times10^{-12}$ erg cm$^{-2}$ s$^{-1}$ in the 1230--1650 \AA\ region, and 
$2.54\times10^{-13}$ erg cm$^{-2}$ s$^{-1}$ in the \ion{C}{4} doublet region
(1545--1555
\AA).  We estimate that the flux calibrations for RU Lupi and T Tau are
accurate to $\sim 15$\%.  Since the telescope breathing was not significant
during the observations of DF Tau and TW Hya, these spectra  
are flux-calibrated with an error of at most 10\%.  
Models of the telescope breating$^1$ accurately predict the presence or
absence of the thermal focus changes for the observations described above.
The low count rates in the  
observations of DG Tau, V836 Tau, V410 Tau, and V819 Tau prevent us from
determining whether they also suffered from the 
same thermal focus variations as
RU Lupi and T Tau.  Based on the telescopic breathing models$^1$ we do
not expect any significant variations in the point spread function during
those observations, and estimate that the flux in those
observations is accurate to better than 15\%.
\footnotetext[1]{see
  http://www-int.stsci.edu/instruments/observatory/focus/ephem.html}

The Doppler correction in the {\it calSTIS} data reduction 
pipeline was in error by a
factor of 1.6$^2$.  We corrected for this problem in the observations of DF
Tau, T Tau, and RU
Lupi by cross-correlating the spectra obtained within the first 1500 s of
each orbit, then subdividing and coadding each 300--500 s interval over 
each observation.  We then applied this correction to the entire
integration for each observation.
This method results in a spectral resolution of about 45,000.  
The low S/N in our observations of DG Tau, V836 Tau, V819 Tau, 
and V410 Tau prevent the application of this method to 
correct for the erroneous wavelength shift.  Those observations have 
$R\sim25,000$.
The $0\farcs5\times0\farcs5$ aperture that we used to 
observe DF Tau and TW Hya is an
unsupported mode of STIS.  By comparing the H$_2$ spectral profiles in our
STIS E140M observations of TW Hya to the STIS E140H ($R\sim100,000$)
observations of TW Hya (Johns-Krull et al., in preparation), we find
that the use of the  $0\farcs5\times0\farcs5$ degrades the resolution of
E140M spectra to $\sim25,000$.
\footnotetext[2]{see http://www.stsci.edu/hst/stis/calibration/pipe\_soft\_hist/update215c.html}

The initial wavelength
calibration was performed using the on-board Pt/Cr-Ne lamps.  We subsequently
re-calibrated the wavelengths by shifting the measured wavelength of the geocoronal
\Lya\ emission line to the predicted location.  The relative wavelength
calibration within an exposure is accurate to $<0.5$ pixels, or $<1.5$ \kms,
and
the absolute calibration across exposures is accurate to $<1$ pixel, or $<3$
\kms\ \citep{Lei01}.  The measured wavelengths of H$_2$ lines from the two DF Tau observations differ by 4
\kms, which is most likely an artifact from our wavelength calibration.  We shift the wavelength
scale in these two spectra by 0.25 pixels 
so that the heliocentric velocities of narrow interstellar \ion{O}{1} 1302 \AA,
\ion{C}{2} 1335 \AA, and \ion{C}{2} 1336 \AA\ absorption lines are equal.

To complement our {\it HST} observations we obtained echelle
spectra of T Tau, DF Tau, V819 Tau, and V410 Tau, with the SOFIN
spectrograph at the {\it Nordic Optical Telescope} ({\it NOT}), covering the entire
optical range at $R=45,000$.
We derive the
projected rotational velocities and heliocentric radial velocities with an
accuracy of $\sim1.5$ \kms\ from these spectra and spectra of template stars.
Because V410 Tau is a spotted, rapidly rotating star with broad
and complex absorption lines, its radial velocity is measured less precisely.
Our velocity measurements are similar to previous estimates \citep[e.g.,][]{Her88}.
 
\section{SOURCES}
Our sample consists of five CTTSs and 3 WTTSs.  The CTTSs were selected
because they are among the brightest CTTSs observed by {\it IUE} and have
a range of disk inclinations, mass accretion rates, and circumstellar
environments.  The three WTTSs were selected to represent varying stages of
disk evolution.  One WTTS, V836 Tau, retains a disk and is weakly
accreting.  Two of the three WTTSs, V410 Tau and V819 Tau, show no
H$_2$ emission and are only briefly discussed.

The known properties for each of our sources, including the radial velocity
($v_r$), the rotational velocity ($v \sin i$), inclination, and
multiplicity are listed in Table 2.  The stellar mass, radius, and
temperatures for the Taurus stars may be found in \citet{Ken95}.  The
properties of TW Hya and RU Lupi are listed in \citet{Web99} and
\citet{Her05}, respectively.  We calculated the accretion luminosity
($L_{acc}$) and mass accretion rates ($\dot{M}_{acc}$) at the time of each
observation using optical and, for DF Tau and TW Hya, NUV spectra, obtained
nearly simultaneously with the FUV observation.  For TW Hya, RU Lupi, T
Tau, and DF Tau, we use extinctions ($A_V$) calculated from the neutral
hydrogen column density in the line of sight to each star, measured from
\ion{H}{1} absorption against \Lya\ emission and H$_2$ absorption detected
in \FUSE\ spectra of the stars \citep[cf.][]{Wal03,Her04}.  If dust grains in our line
of sight to these sources are larger than the average interstellar grain,
then $A_V$ would be underestimated by $R_V/3.1$, where $R_V$ is the
total-to-selective extinction with an average interstellar value of 3.1 \citep{Car89}.

\subsection{T Tau}
As the archetype of the entire class of stars, T Tau is one of the most
studied CTTSs.  \citet{Wal03} and \citet{Bec04} present an overview of T Tau, which is
comprised of the optically bright T Tau N separated by $0\farcs7$
from the IR companion T Tau Sab.

T Tau N contributes all of the flux in the FUV because T Tau Sab is
heavily obscured \citep{Kor97,Duc05}.  Most previous estimates of the
extinction to T Tau N are 
$A_V\sim1.5$ mag \citep{Ken95,Gul00,Whi01}.  \citet{Cal04} estimated
$A_V=1.8$ mag by fitting the accretion continuum and photospheric emission
from T Tau, although the anomalously complex NUV continuum is
poorly fit with standard accretion continuum models.  An extinction
of $A_V\sim1.5$ mag would severely attenuate FUV emission, but emission in
the \ion{C}{3} 977 \AA\ and 
\ion{O}{6} 1032 \AA\ lines  was detected by {\it FUSE} \citep{Wil02}.  
\citet{Wal03} 
used the total hydrogen column density in the line of sight to T Tau N to
calculate $A_V=0.3$ mag, which we adopt here.

Various studies suggest an inclination of the disk axis from our line of sight 
of $8-23^\circ$ from our line of sight \citep{Her86,Her97,Eis98}.
IR H$_2$ emission imaged by \citet{Van94} and \citet{Her97}, 
and long-slit \HST/STIS FUV spectra obtained by \citet{Wal03} 
and \citet{Sau03} show spatially extended H$_2$ emission.
Images of shock tracers such as [\ion{Fe}{2}] and
the extended H$_2$ emission reveal an extensive network of nebulosity, 
including Burnham's nebula and HH 155, produced by outflows 
interacting with ambient molecular material 
\citep[e.g.,][]{Her96,Her97,Sta98,Sol99}.

The nominal distance to T Tau and to the other stars in 
the Taurus molecular cloud is $\sim140$ pc.  \citet{Loi05} 
calculated a distance of $141\pm2.8$ pc to T Tau by measuring 
the parallax in high precision astrometric observations of non-thermal
radio emission from T Tau S.

\subsection{DF Tau}
DF Tau is a binary system with a separation of $0\farcs09$ and a position
angle of $\sim270^\circ$ at the time of our 
observations \citep{Whi01,Har03,Scha03,Har04}.  
\citet{Scha03} used the decreasing position angle of the pair ($300^\circ$
in 1994 to $262^\circ$ in 2002) to study the orbital motion of the stars.
They suggest that the optically bright component, DF Tau
A, is the dimmer component in the near-IR.  
\citet{Scha03} also find
that the V-band emission from the primary star varies by 1.5 mag, while the 
secondary varies by less than 0.2 mag.  They find that the two stars
have a similar temperature, 
mass, and luminosity, but mass accretion rates of $\sim10^{-8}$ 
and $\sim10^{-9}$ \mdotyr\ for DF Tau A
and DF Tau B, respectively.  As a result, DF Tau A dominates the U-band emission.  
Although these characteristics are all
consistent with DF Tau A having a higher mass accretion rate than DF Tau B, 
they could also be explained by invoking
a larger extinction to DF Tau B.  \citet{Fur05} do not find any evidence
for disk evolution, either by dust settling or a clearing of
the inner disk, from {\it Spitzer} IRS spectra of DF Tau.

The disk is inclined to our line of sight by 60--85$^\circ$, based on measurements of $v\sin i$
and a rotation period of 8.5 d \citep{Har89,Joh01}.
Previous extinction estimates range 
between $A_V=0.15$ and $0.45$
mag. \citep{Ken95,Gul98,Whi01}.  Following \citet{Her04}, we 
measure $\log N$(\ion{H}{1})=$20.75$ 
from the absorption against the red side of the \Lya\ emission line.  
We also estimate $\log N$(H$_2)=20.2^{+0.3}_{-0.5}$ from 
the H$_2$ absoprtion against the \ion{O}{6} 
and \ion{C}{3} emission lines in 
the {\it FUSE} spectrum of DF Tau.  The uncertainty in this 
measurement is large because the measurement is inferred 
indirectly from the
weak emission in the \ion{O}{6} 1038 \AA\ line relative 
to the emission in the \ion{O}{6} 1032 \AA\ line.  The H$_2$ 
excitation temperature is also uncertain.  The total hydrogen column density 
$\log N$(H)$\sim20.95$ corresponds to $A_V=0.5^{+0.15}_{-0.2}$ 
mag assuming the standard interstellar gas-to-dust ratio \citep{Boh78}.
Our line of sight to the star may intercept the flared disk, which could have dust grains
larger than is typical for the ISM.  Any grain growth in our line of sight,
either in a disk or the Taurus molecular cloud,
will increase the gas-to-extinction ratio and cause us to
underestimate the extinction.  On the other hand, the flared disk may be
deficient in grains if they have settled, relative to the gaseous disk.

In the long-pass optical acquisition image obtained prior to our 
small-aperture STIS observation of DF Tau,
the pair are resolved with a separation of 96 mas and a position angle of
266$^\circ$,
with DF Tau A being about 13\% brighter than DF Tau B in this band.  This
separation and position angle are consistent with the relative positions of
DF Tau A and B measured by \citet{Scha03} for the same epoch.  The peak-up
image and the echelle spectra, both obtained with the
$0\farcs2\times0\farcs06$ aperture, have position angles offset by
$\sim48^\circ$ from the dispersion direction, and only include one star.
The peak-up presumably found DF Tau A, the brighter of the pair in the CCD.

Only one star is apparant in the
acquisition [\ion{O}{2}] image prior to the large-aperture observation of DF Tau.  
 Both stars are included in our $0\farcs5\times0\farcs5$ observation 
of DF Tau with a position angle offset by 51$^\circ$ from the dispersion 
direction.  The pair may not be resolvable and the secondary is not
detected in the FUV.  
Assuming that DF Tau A dominates emission in the [\ion{O}{2}] filter and 
in the FUV, then the fainter DF Tau B is located at +6 \kms\ in the 
dispersion direction.  Because the two observations were obtained 
with a PA that differed by 180$^\circ$, any spectroscopically resolveable
emission contributed by DF Tau B  
would appear slightly blueshifted in one observation and redshifted in the other 
observation.

We estimate a mass accretion rate of about $3\times10^{-8}$ \mdotyr\ based
on the NUV-optical spectrum obtained just before our observation of DF Tau
with the large aperture.  During our small aperture observation,
we measured an accretion luminosity about 15 times lower based on the
emission in the G430L spectra shortward of the Balmer jump.
The optical emission longward of the Balmer jump
was about 5 times fainter during the small aperture observation than in the 
large aperture observation.  The $N$(\ion{H}{1}) and therefore the extinction 
does not change between the two observations.

The FUV spectrum obtained with the smaller aperture is
about three times fainter than that obtained with the larger aperture, 
either because of variability in the mass accretion rate, the different
aperture size, or the pointing
being slightly offset from DF Tau A.  These two observations, obtained with 
different apertures at different epochs, are analyzed separately.   

\subsection{RU Lupi}
RU Lupi was discussed in detail by \citet{Gio95} and \citet{Her05}.  
Although most previous studies have assumed that RU
Lupi is a single star, \citet{Gah04} detected periodic radial velocity
changes of 2.5 \kms\ in photospheric absorption lines, which may indicate
the presence of a spectroscopic companion.  Indirect evidence indicates
that the disk and magnetosphere of RU Lupi are probably observed close to
pole-on \citep{Gio95,Her05}.  \citet{Gah05} derive an inclination of
$\sim23^\circ$ based on the $3.7$ d period and $v\sin i=9.0$ \kms.

RU Lupi is one of the most heavily veiled CTTSs with a variable H$\alpha$
equivalent width that peaks at 210 \AA.  \citet{Her05} described the {\it
HST}/STIS observations of RU Lupi analyzed here.
The mass accretion rate onto RU Lupi of $3\times10^{-8}$ at the time of our
observation is 
high for a K7 CTTS.  The outflow from RU Lupi is also very strong, as seen in P Cygni line
profiles of neutral, singly-ionized, and doubly-ionized species.  The extinction to RU Lupi is $A_V\sim0.07$ mag, based on the
$\log N$(H)$=20.1$ to the star \citep{Her05}.  Like Taurus, the
Lupus molecular cloud is probably located at 140 pc \citep{deZ99,Ber99}.

\subsection{DG Tau}
DG Tau is a binary CTTS characterized by strong accretion and a powerful
bipolar outflow.  {\it HST}/WFPC2 images of DG Tau reveal an edge-on disk 
that obscures the star \citep[]{Kri95}.  The well-studied jets from DG Tau have
provided powerful insights into the production of slow and fast winds from
CTTSs \citep[e.g.,][]{Bac02,And03,Har04}.
The jet has a position angle of 226$^\circ$ and an inclination to our line
of sight of 38$^\circ$ \citep{Bac02}.  In \S 5 we argue that the 
FUV H$_2$ emission from DG Tau is produced in an outflow.  
The FUV spectrum includes very little emission in lines other than H$_2$.  
The long-slit optical spectrum obtained with the 52$^{\prime\prime}\times 0\farcs2$ aperture 
included the star.  The star was acquired with a long-pass optical filter 
and the optical peak-up succeeded in centering the optical emission.  
 At least some of the X-ray emission from DG Tau is extended beyond the
 star and likely produced by outflows \citep{Gue05}. 
The FUV emission from DG Tau may also be significantly extended beyond the
star.  We estimate $L_{acc}=0.22$ $L_\odot$, which corresponds to
$\dot{M}\sim3\times10^{-8}$ \mdotyr.

We adopt an extinction of $A_V=1.6$ mag to DG Tau, based on models of the 
NUV and U-band continuum by 
\citet{Gul00}.  However, this extinction may apply only to the star itself 
but not to the warm molecular gas, because the star is viewed through its edge-on disk. 

\subsection{TW Hya}
The namesake of the sparsely populated TW Hya association \citep{Kas97,Web99}, TW
Hya is the closest (56 pc) and UV-brightest known CTTS.  Even though 
it is $\sim 10$ Myr old \citep{Web99},
TW Hya is still weakly accreting, with $\dot{M}_{acc}=2\times10^{-9}$ $M_\odot$ yr$^{-1}$ at the
time of our observations \citep{Her04}.  
Since TW Hya is isolated from molecular 
clouds that are 
typically associated with CTTSs, the extinction is
negligible \citep{Her04}.  
Images of the disk reveal that it is observed nearly face-on
\citep{Wil00,Kri00,Tri01,Wei02}.   
\citet{Qi04}
used submillimeter observations of the CO $J=3-2$ and $J=2-1$ 
lines to measure an inclination of $7\pm1^\circ$.
From a lack of near-IR excess emission from the disk, \citet{Cal02} inferred
that the warm micron-sized dust in the disk within 4 AU of the central star
is optically thin. 

The FUV spectrum of TW Hya was described by
\citet{Her02} and the fluorescent H$_2$ emission was modelled by
\citet{Her04}. \citet{Wei00} measured a flux of $1\times10^{-15}$ \erg\ in
the rovibrational 1-0 S(1) line at 2.1218 $\mu$m, and \citet{Ret04}
detected CO emission from TW Hya.

\subsection{V819 Tau}
V819 Tau shows no significant excess continuum emission at $\lambda<12$
$\mu$m but exhibits strong excess emission at 12, 25, and 60 $\mu$m,
which suggests the presence of a cold disk with a central dust hole
\citep[cf.][]{Skr90,Wol96}.  \citet{Bar03} placed a flux upper limit of
$3.0\times10^{-15}$ \erg\ in the 1-0 S(1) line.  The extinction of
$A_V\sim1.5$ mag to V819 Tau 
\citep{Ken95,Whi01} strongly attenuates any FUV emission.  We do not discuss this
star further.

\subsection{V836 Tau}
V836 Tau is typically classified as a WTTS, although it has a near-IR
excess, indicating the presence of a disk
\citep{Mun83,Skr90,Ski91,Ken95,Wol96}.  V836 Tau also has H$\alpha$
emission with a variable equivalent width of $9-25$ \AA\
\citep{Skr90,Har95,Whi04} and an inverse P-Cygni profile \citep{Wol96},
both indicative of accretion.
We adopt $A_V\sim0.6$ mag to V836 Tau \citep{Ken95}.
\citet{Whi04} estimate a mass accretion rate of $\sim10^{-8}$ \mdotyr.  At
the time of our observation, we estimate a mass accretion rate of
$\sim10^{-9}$ \mdotyr\ based on the measured \ion{C}{4} flux of
$1.4\times10^{-15}$ erg cm$^{-2}$ s$^{-1}$ and the relationship between
$\dot{M}$ and \ion{C}{4} flux calculated by \citet{Joh00}.

\subsection{V410 Tau}
V410 Tau is a system with three WTTSs that are coronally active and have no
detectable IR excess \citep{Ken95}.  
\citet{Whi01} measured a separation of $0\farcs07$ between the A and B
components with an uncertain position angle, and a
separation of $0\farcs29$ between the A and C components with a position angle of
359$^\circ$.  Since they found that the primary dominates the U-band
emission, we expect that it also dominates the FUV flux.  Our FUV
observation most likely includes both the A and B components but does not
include V410 Tau C.

\citet{Ken95} estimated $A_V=0.03$ mag, while \citet{Whi01} calculated
$A_V=0.67$ mag.  Although this disparity is large, it does not
significantly impact our analysis.
Strong variability in the emission from V410 Tau has been well studied at
radio to X-ray wavelengths \citep[e.g.,][]{Ste03,Fer04}.  \citet{Ryd83}
first reported a 1.87 d period that has been attributed to
stellar spots.

\section{CHARACTERIZING THE H$_2$ EMISSION}
Strong ($A_{ul}\sim10^{8}$ s$^{-1}$) electronic transitions of H$_2$ occur
throughout the FUV wavelength range.  Cold H$_2$ can be excited by photons
at $\lambda<1120$ \AA, while warmer H$_2$ can also be excited by photons at
longer wavelengths.  Lyman-band (B-X) transitions tend to occur at longer
wavelengths than Werner-band (C-X) transitions because the B electronic
state has a lower energy than the C electronic state.  Once electronically excited, 
the H$_2$ molecule almost immediately decays to the ground (X) 
electronic state.
The radiative decay from the B electronic state will occur by one of many
different transitions that have similar branching ratios.  Typically only a
few transitions from the C electronic state have large branching ratios.
The H$_2$ molecule in the B or C electronic state will dissociate by
radiative decay to the ground vibrational continuum between 0--50\% of the
time, depending on the energy of the upper level.

Warm H$_2$ gas can absorb photons throughout the FUV.
Therefore, a flat radiation field would excite a myriad of upper levels and
produce a pseudo-continuum of densely-packed H$_2$ lines.  
However, the FUV emission from  CTTSs is dominated by emission 
lines, in particular
\Lya.  Consequently, most of the detected H$_2$ lines are from 
one of the 10--25 different upper levels that are excited by transitions with wavelengths 
coincident with \Lya\ emission.
\Lya-pumped H$_2$
fluorescent emission appears to be common to all CTTSs.  The \Lya\ excitation and
decay process for CTTSs is 
described in \citet{Her02}.  All lines discussed in this paper are
Lyman-band lines, except where noted.  The set of emission lines from a
single upper level is termed a {\it progression}, and is referred to by the
vibrational and rotational quantum number of the upper level, ($v^\prime,J^\prime$).
The amount of emission in a progression
depends on the population of H$_2$ in the lower level of the transition, 
the oscillator
strength of the pumping transition, and the radiation field at the
transition wavelength.  
Table 3 lists the pumping transitions for the detected emission lines, together
with the oscillator strength $f$, the energy $E^{\prime\prime}$ of the
lower level of the pumping transition, 
the theoretical dissociation percentage $P_{dis}$ from the upper level
calculated by \citet{Abg00}, and the velocity of the pumping transition
from
\Lya\ line center.  
Table 4 lists the number of lines and the total observed flux from each
upper level for every source.  

Figure 2 shows the observed \Lya\ emission and locations of the pumping
transitions for each source.  For every source discussed here except TW Hya, most or all
of the intrinsic \Lya\ emission is attenuated by \ion{H}{1} in the
interstellar medium and 
stellar outflows.  We do not detect any \Lya\ emission from T
Tau, DG Tau, V836 Tau, or V410 Tau.   We detect only weak \Lya\ emission
located far from line center from DF Tau and RU Lupi.  Only for TW Hya do
we detect strong \Lya\ emission both longward and shortward of \Lya\ line
center, because TW Hya is a weak accretor and isolated from molecular
clouds that are typically associated with young stars.
The strength and large width of the \Lya\ emission 
from TW Hya leads to a more extensive network of observed H$_2$ 
lines than from the other sources in our sample.  

The observation 
of DF Tau obtained with the small aperture has lower S/N than that 
obtained with the large aperture, and shows fewer lines as a result.  
No significant differences in H$_2$ emission are detected between 
those two observations.

Several H$_2$ lines pumped by \ion{C}{4} from highly excited rovibrational
levels in the ground electronic state are discussed in \S4.3.

\subsection{Identifying H$_2$ lines}
The FUV spectra of the CTTSs TW Hya, DF Tau, RU Lupi, T Tau, and DG Tau, and
the WTTS V836 Tau include many narrow H$_2$ emission lines.  We do not detect
any H$_2$ emission in the spectra of V819 Tau or V410 Tau.
We identify H$_2$ lines using the linelist of \citet{Abg93}.  
Most of the lines are Lyman-band lines pumped by \Lya, because Werner-band
lines pumped by \Lya\ are brightest at 
$\lambda<1200$ \AA, where the sensitivity of STIS is low.
If one line from an upper level
is present, then several other lines with large branching ratios from the
same upper level should also be 
present.  In order to positively identify a line as H$_2$, we require that 
several lines from a 
given upper level be present with relative fluxes consistent with
branching ratios.

Based on models constructed for the H$_2$ fluorescence
detected in the TW Hya spectrum (Herczeg et al. 2004, see also \S 4.2), 
we use observed fluxes from one or a few lines from an upper 
level to predict fluxes for all the lines
in that progression.
These models allow us to identify many weak emission features as
H$_2$ lines.  
Figure 3 shows the observed and model fluxes of sixteen lines originating 
from (\Vup,\Jup)=(2,12) in the
DF Tau spectrum.  We use the fluxes in the strongest lines in 
that progression, such as 2-8 P(13) at 1588.7 \AA\ and 2-5
P(13) at 1434.5 \AA, to identify and predict the
fluxes of the weaker H$_2$ lines, such as 2-1 R(11) at 1185.2 \AA\ and 2-3
P(13) at 1325.3 \AA. 
These models provide a rigorous check on questionable line 
identifications because they predict the fluxes of all other
possible lines from the same upper level.

In rare cases, we identify only one or two
lines from a single upper level as Lyman-band H$_2$ lines when the
following conditions are met: (i) the lines have an 
appropriate width and velocity shift, (ii) all other lines from the same upper
level are either too weak to be detected or are obscured by wind absorption 
or other
emission lines, and (iii) we expect emission in that progression based 
on the observed emission at wavelengths of possible pumping transitions.
Generally only a few
Werner-band lines from a single upper level have large branching ratios, and these lines
preferentially occur at $\lambda<1185$ \AA, where \STIS\ has low
sensitivity.  Therefore, we relax the requirement of detecting several
lines from a single upper level for the identification of  Werner-band
H$_2$ lines.

H$_2$ emission lines were previously identified in the STIS spectra of TW
Hya \citep{Her02} and RU 
Lupi \citep{Her05}.   In this paper, we identify H$_2$ lines from V836 Tau, T
Tau, DG Tau, and two spectra of DF Tau, and expand the list of H$_2$ lines
observed from TW Hya.  
Table 5 lists line identifications, fluxes, and branching ratios of
observed Lyman-band H$_2$ 
emission lines that are pumped by \Lya\ from each star, sorted by progression.
Table 6 lists the same properties for highly-excited H$_2$ emission lines
that are pumped by \ion{C}{4}.  Table 7 
lists the blended H$_2$ lines.  In some spectra, lines such as 
the 1455 \AA\ blend are resolved into two separate lines, while 
in other spectra we were unable to discriminate between the two 
lines.  Table 8 lists the Werner-band H$_2$ lines, most of which
are also pumped
by \Lya, although one Werner-band H$_2$ line from RU Lupi is pumped by  
strong emission in the \ion{O}{6} 1031 \AA\ line. 
Outside of regions with strong lines such as \ion{C}{4} or \ion{O}{1}, 
the H$_2$ linelist for TW Hya is complete for $\lambda>1200$ \AA\ down 
to a flux level of $\sim4\times10^{-15}$  erg cm$^{-2}$ s$^{-1}$, while 
the H$_2$ linelists for RU Lupi and DF Tau are complete down to $\sim1\times10^{-15}$ erg 
cm$^{-2}$ s$^{-1}$, and that of DG Tau and V836 Tau are complete down 
to $\sim4\times10^{-16}$ erg cm$^{-2}$ s$^{-1}$.
The H$_2$ lines from T Tau are more difficult to identify because they are
broad, which lowers the S/N in each pixel, and are weak
relative to many other strong lines in the spectrum, which increases the
likelihood of masking by other emission or wind absorption lines.  As a
result, our H$_2$ linelist for T Tau may not be complete.

Figures 4--7 present four spectral regions with strong H$_2$ lines.
The strongest H$_2$ lines are transitions from the upper 
levels (\Vup,\Jup)=(1,4),
(1,7), (0,1), (0,2), and (2,12).  These lines are all pumped by 
transitions located between 0 and 600 \kms\ from \Lya\ line center (1215.67
\AA).  The pumping transitions typically have large oscillator strengths
and H$_2$ lower levels with relatively low energies.
The locations of the pumping transitions agree with results from
\citet{Ard02a}, who detected only H$_2$ lines pumped on the red side of
\Lya\  in GHRS spectra of several CTTSs.
TW Hya, RU Lupi, DF Tau, DG Tau, and V836 Tau also exhibit weaker H$_2$
emission pumped shortward of \Lya\ line
center.  Many possible pumping transitions are present shortward of \Lya,
and are all detected in the TW Hya spectrum.
In the spectrum of DF Tau, we detect H$_2$ emission from only two levels
that  are pumped shortward 
of \Lya, and both are pumped at $v<-300$ \kms\ relative to \Lya\ line center.
However, some H$_2$ lines from DF Tau are pumped at $v=+1100$.
 Therefore, the H$_2$ lines from DF 
Tau and T Tau indicate that the centroid of the \Lya\ emission that
irradiates the warm molecular gas is longward of line center.

As described above, the \Lya\ profile seen by the warm molecular gas differs
from the observed \Lya\ emission because of \ion{H}{1} attenuation in
our line of sight to the star.  Some \Lya\ emission may be attenuated
by \ion{H}{1} prior to irradiating the warm molecular gas.  In principle,
the H$_2$ gas may also see a different \Lya\ profile if geometrical effects
result in anisotropic \Lya\ emission.
The observed \Lya\ emission from RU Lupi, TW Hya, and DF Tau extends to
1219 \AA, 1220 \AA, and 1221 \AA, respectively.  Table 4 shows that the H$_2$
emission lines from DF Tau, but not from RU Lupi or TW Hya, are pumped at
$\lambda>1220$ \AA.  The H$_2$ lines pumped at $\lambda>1218$ \AA\ are also
much stronger from TW Hya and DF Tau than from RU Lupi.  DF Tau is viewed
edge-on by both the observer and the H$_2$, but TW Hya and RU Lupi are
observed face-on by the observer and edge-on by the H$_2$.  Therefore, the
presence or absence of the red wing in the \Lya\ emission profile appears
to be independent of viewing angle.

The H$_2$ lines pumped at large velocities from \Lya\ line center are not
detected in {\it IUE} spectra of HH 43 and HH 47
\citep{Sch83,Cur95}, or emission extended by up to
20$^{\prime\prime}$ from T Tau \citep{Wal03,Sau03}.  Those spectra show
H$_2$ emission from only the upper levels (1,4) and (1,7) that are pumped at +14 and +99 \kms\ from \Lya\ line center.    The accretion processes associated with CTTSs produce much
broader \Lya\ emission than is produced by shocks due to interactions between
outflows and molecular clouds.  Thus the pattern of H$_2$ fluorescence is a good diagnostic of whether accretion processes or shocks in the interstellar medium provide the \Lya\ pumping photons.

\subsection{H$_2$ line fluxes}
We fitted the H$_2$ lines with Gaussian profiles to measure the central
wavelength, width, and flux.  The H$_2$ lines from T Tau and RU Lupi both show
significant asymmetric blueshifted emission.  We fit each line from those two sources
with one narrow, bright Gaussian component and a fainter broad component that is
blueshifted.  The width, velocity shift, and percent of the total flux in
the weaker component were fixed based on fits to coadded emission lines (see
\S 5.1). 
The H$_2$ lines from DF Tau also show a weak blueshifted asymmetry.  By
fitting single Gaussians to all of the lines from DF Tau, we may be
underestimating the true flux by $\sim 5$\%.
This fitting process assumes that the line profile does not depend on the upper level.  This method is appropriate for RU Lupi but may result in underestimating the flux in 
lines originating from \Vup,\Jup=(1,4) and (1,7) from T Tau and DF Tau (see \S 5.1).
Tables 5--8 list the H$_2$ lines from each star, sorted by upper level.

\citet{Woo02} and \citet{Her04} constructed Monte Carlo  
models of H$_2$ fluorescence in a plane-parallel slab by computing on the optical depths of the various lines using the branching 
ratios calculated by \citet{Abg93}. 
Depending on the column density and excitation
temperature of the H$_2$ emission region, large line opacities can occur
when a low-energy lower level is heavily populated.
Transitions from lower
levels with large energies remain optically thin because those lower levels
have negligible
populations.  This effect tends to weaken the lines at short wavelengths.
These models have been used to probe the
temperature $T$ and column density $N$(H$_2$) of the H$_2$ emission region.
In this paper, we use the model with the best-fit parameters of $T=2500$ K
and $\log N$(H$_2$)=18.5 that \citet{Her04} found for the H$_2$ emission
from TW Hya.  We apply results from that model here to check our line identifications (\S 4.1) and to
estimate fluxes in the undetected lines.  

We scale the relative line fluxes predicted by the model to match the
observed fluxes, after correcting for extinction.  The
predicted line fluxes are somewhat uncertain because the physical conditions and the geometry of the warm molecular gas could be different for stars with different emission sources (see \S 5.4).  Nonetheless, these models explain most of the observed line fluxes successfully.  
 Figures 3--7 compare the observed flux (solid line in Fig. 2 and shaded
 regions in Figs. 4--7) with the model flux (dashed lines).
Tables 5--6 list estimates of the total emission from an
upper level by correcting for lines that are unseen because they are
outside of our wavelength range, are masked by strong emission lines,
attenuated by wind absorption, or are too weak to be detected.  

Although the relative line fluxes are weakly
sensitive to extinction, we are unable to improve
upon the existing extinction estimates because of the low S/N for 
lines at short wavelengths.  The relative line fluxes
are consistent with our adopted extinctions in all cases.  The H$_2$ lines in 
the T Tau spectrum 
indicate $A_V<1.0$ mag, which suggests that the lower extinction value toward
T Tau of $A_V=0.3$ mag adopted here is appropriate.

The observed H$_2$ fluxes from DF Tau are 2.5--3 times lower in our
observation with the small aperture compared to the observation with the
large aperture, but the \Lya\
emission was only 1.73 times smaller.  Since H$_2$ is pumped by \Lya, 
the observed H$_2$ fluxes
should be directly proportional to the strength of \Lya\ emission.  However,
we cannot draw any significant conclusions from this discrepancy because 
the large aperture contains both stars, while the $0\farcs2\times0\farcs06$ 
aperture contained only one star.

Table 9 lists the observed flux, not corrected for extinction, in strong 
FUV emission lines and in the strongest
five H$_2$ progressions.  The typical ratio of observed C IV to
H$_2$ flux
in the strongest progression is about 7:1.  We therefore expect that
between two to five  
progressions from the WTTS V410 Tau should have fluxes $>4\times10^{-15}$
erg cm$^{-2}$ s$^{-1}$.  We place flux upper limits of $1-2\times 10^{-15}$
erg cm$^{-2}$ s$^{-1}$ in each
progression from V410 Tau by finding flux upper limits for individual
emission lines and then applying the models described above to calculate an 
upper flux limit in the entire progression.  These upper limits are about a
factor of two below  
the H$_2$ flux we crudely expect from the previously described correlation
with \ion{C}{4} emission.  We conclude that 
a significant reservoir of warm H$_2$ is probably not present around V410
Tau.  Since a large extinction attenuates almost all of the FUV emission from the WTTS V819 Tau, the non-detection of H$_2$ from this star is not significant.

\subsection{Highly Excited H$_2$}

Most of the detected H$_2$ lines are photoexcited from lower
levels in the ground
electronic state with energies of $1-2$ eV.  These levels are populated sufficiently with excitation temperatures of
$T=2000-3000$ K to explain the observed emission \citep{Bla87,Woo02,Her04}.
\citet{Her02} detected emission from the (\Vup,\Jup)=(0,17) and (0,24) levels from TW Hya,
yet these upper levels cannot be excited by \Lya.  We confirm the presence
of these lines in the spectra
of both TW Hya and DF Tau.  We also identify emission in lines from (1,14)
in the spectrum of TW Hya.  None of the transitions that could pump these
levels are coincident with \Lya.   
These upper levels may be excited by 
 \ion{C}{4} via 0-3 P(25) 1547.97 \AA, 0-5 P(18) 1548.15 \AA, and 1-7 R(13),
respectively.  The \ion{C}{4} 1548 \AA\ resonance line is typically the
second strongest line in FUV spectra of CTTSs, after \Lya, and the three
possible pumping  
transitions have strong oscillator strengths.  However, the lower levels
(\Vlo, \Jlo)=(3,25), (5,18), and (7,13)
have energies of 4.2, 3.8, and 3.8 eV,
respectively, and cannot be thermally populated  
at temperatures where H$_2$ is present.  These lower levels are also not
directly populated by the fluorescence and subsequent decay into
vibrationally excited levels of the ground electronic state.  The
fluorescence and subsequent decay increases the 
vibrational excitation without substantially changing the rotational
excitation of H$_2$.  Excitation of H$_2$ by FUV pumping and fluorescence therefore cannot explain the high rotational excitation seen here.   Some non-thermal process, perhaps
involving the  
formation of H$_2$, may also populate these highly energetic levels.  

We investigated the excitation conditions required to produce this  
emission by forward modelling the H$_2$ spectrum.  We constructed synthetic
H$_2$ spectra by modelling a isothermal, plane parallel slab of H$_2$ 
following the procedure described by \citet{Woo02} and \citet{Her04}.   The 
amount of emission absorbed by an H$_2$ transition, $F_{H2}$, is given by:
\begin{equation}
F_{H2}=\eta F_{pump}\int_\lambda [1-e^{-\tau_\lambda(T,N(H_2))}]{\rm d}\lambda,
\end{equation}
where $F_{pump}$ is the flux at the pumping wavelength, $\eta$ is the solid
angle filling factor of  
H$_2$ as seen from the \Lya\ emission region, assumed to be 0.25 \citep{Her04}, $T$ and
$N$(H$_2$) are the temperature  
and column density of the slab, and the integral is the effective
equivalent width of the transition.   The entire FUV spectrum is required
to calculate the excitation of every upper level of H$_2$.  
We used the {STIS} spectrum of TW Hya to estimate the flux at
$\lambda>1187$ \AA, and a {\it FUSE}  
spectrum of TW Hya to estimate the flux at $\lambda<1187$ \AA\ (Herczeg et
al., in preparation). 
Among the many models run, we present results for synthetic spectra based
on an H$_2$ layer with 
$\log N$(H$_2$)$=18.5$ (units cm$^{-2}$) and $T=2500$ K (our standard model), and that standard model with 
an additional layer of $\log N$(H$_2$)$=17.0$ and
$T=2.5\times10^4$ K. 
Differences between the two model spectra are attributed to the small
reservoir of highly excited gas.  We also calculate 
spectra for models of gas that is irradiated by all FUV photons except for
\Lya.  Figure 8 compares the observed TW Hya spectrum to three of these models.

The H$_2$ lines from the (0,17), (0,24), and (1,14) upper levels are produced by photoexcitation by \ion{C}{4} in the highly excited layer of H$_2$.  Lines in these three progressions are the strongest Lyman-band
lines that are not pumped by \Lya.
 Based on
these models, we find many other weak features in the spectrum that could
also be attributed to pumping by \ion{C}{4}, \ion{C}{2}, and several other
strong emission lines.  While the temperature and column density for this hot layer
are poorly constrained at present, these results demonstrate the need
for a thin layer of highly-excited H$_2$ to explain these lines.

With the presence of the highly excited H$_2$ layer, our models 
predict that many undetected H$_2$ lines should be much stronger
 than the weak lines from (0,17), (0,24), and (1,14).
 Since all of these undetected lines should be pumped by \Lya,  
we are forced to
 conclude that somehow \Lya\ photons are prevented from irradiating the bulk of the highly-excited H$_2$ gas.
 This observational result is unexpected and not easily explained.  
We therefore suggest a highly speculative possible explanation, although there may be other more physical explanations.
In
 principle, the
 highly-excited H$_2$ could be mixed in with both the warm ($\sim 2500$ K)
 H$_2$ layer and colder gas below.  Scattering by
 \ion{H}{1} in the warm surface layer could then increase the path length of \Lya\
 photons,
 leading to preferential attenuation of \Lya\ emission by dust.  In this
 case, the FUV radiation field that irradiates the gas beneath the thin
 surface layer may be dominated by emission in lines such as \ion{C}{4}.
 We caution the reader that this speculative scenario is outlined here only to
 describe the a process that could irradiate a surface with \ion{C}{4}
but not \Lya\ emission.

 The highly-excited H$_2$ emission is detected from two sources, TW Hya and
DF Tau, that have strong
FUV continuum emission, but is not detected from RU Lupi, which does not
have a strong FUV continuum.
Since the FUV continuum is likely produced by electron excitation of H$_2$
\citep{Ber04}, it could be
related to the highly-excited H$_2$.  Electronic excitation of H$_2$, like FUV pumping, will result in significant vibrational excitation but not significant rotational excitation..
However, if the H$_2$ formation rate is sufficiently large, it could
produce a population of H$_2$ in
highly excited rovibrational levels.  H$_2$ may form by many routes, including on grains in gas at $T<500$ K, by associative detachment of $H^-$, or by dissociative recombination of H$_3^+$.  These and other H$_2$ formation processes produces a population of highly excited H$_2$ \citep[e.g.][]{Tak99,Bie79,Kok01}.


\subsection{Variability of H$_2$ emission from T Tau}
\citet{Wal03} detected emission in the red wing of the \Lya\ line in two of three 
long-slit G140L spectra of T Tau obtained in our program, but 
no \Lya\ emission in the echelle spectrum of T Tau.  Only this red wing is
seen because 
the core is absorbed by interstellar \ion{H}{1}, and any 
blueshifted emission is absorbed in the stellar winds.
The non-detection of any \Lya\ emission in two of the four spectra is the
result of either a variable \Lya\ line width or a variable \ion{H}{1}
column density in our line of sight.
\citet{Wal03} conclude that these changes are not caused 
by a different \ion{H}{1} column density in our line of sight because 
the shape of the \Lya\ profile did not change, even though the 
detected \Lya\ line strength was different in the two observations.
We confirm this result by relating the presence of emission in the red wing
of \Lya\ to the relative flux in each progression.

Figure 9 shows the \Lya\ and 1400--1500 \AA\ regions for 
our three long-slit spectra and our echelle spectrum, convolved 
to the spectral resolution of G140L ($R\sim1000$).  We scale the 
observed emission to equal the flux in every observation between 
1495--1510 \AA, a region dominated by several strong H$_2$ lines 
from (1,4) and (1,7).
Based on the strength of the extended H$_2$ emission, at most 10\% of
the on-source emission in H$_2$ lines from (1,4) and (1,7)
is related to the extended shocks seen in long-slit spectra.
The relative flux of H$_2$ emission lines from the same upper level 
also remains constant during the four observations.  A higher 
extinction in an observation would suppress lines at shorter wavelengths. 
We conclude that the extinction to T Tau remains constant during these
observations.

When the redshifted \Lya\ emission disappears,  
the relative H$_2$ emission in lines from (0,1), (0,2), and (2,12) 
decreases substantially, relative to that from (1,4) and (1,7).  The
echelle observation shows the strongest emission in hot accretion lines
such as \ion{C}{4}, H$_2$ lines from (1,4) and (1,7), and the continuum,
but has no detectable redshifted \Lya\ emission and as a result shows the
weakest absolute flux in H$_2$ lines from (0,1), (0,2), and (2,12).
Since the lines from (0,1), (0,2), and (2,12) are pumped 
at $+379$, $+487$, and $+551$ \kms, respectively, from \Lya\ line 
center, the strength of these lines probe the strength of the red 
wing of \Lya.  The lines from (1,4) and (1,7), which are pumped 
at $+99$ and $+14$ \kms, depend on the emission near \Lya\ line 
center.  The changes in the intrinsic \Lya\ profile are seen both directly
in our observations and by the H$_2$ gas, even though we are viewing the
system face-on while the H$_2$ is likely viewing the star with a large inclination.

\section{CONSTRAINTS ON THE ORIGIN OF THE H$_2$ EMISSION}
In this section, we analyze the spectral and spatial properties of the
H$_2$ emission to identify the source of the H$_2$ emission.  Strong lines
within each progression are coadded for each star to improve the S/N in the
spectral profiles.  We consider only lines in the five strongest
progressions, from (\Vup,\Jup)=(1,4), (1,7), (0,1),
(0,2), and (2,12).
  
We group the lines from (1,4) and (1,7) together
because these progressions are pumped at $+99$ and $+14$ \kms\ from \Lya\
line center,  and consequently may be excited by a narrow \Lya\ emission line.
 The progressions from (0,1), (0,2), and (2,12) are grouped together
 because they are  pumped by transitions between 380 to 550 \kms\ from
 \Lya\ line center, which requires broad \Lya\ emission.

\subsection{Spectral Profiles of the H$_2$ Emission}
Figure 10 shows the normalized spectral profiles of the coadded H$_2$ emission
lines from (0,1), (0,2), and (2,12).  
Table 10 lists the parameters of Gaussian fits to these profiles, including
the velocity 
relative to the star, the intrinsic FWHM deconvolved from the instrumental
broadening, and the percent of 
flux in each component.  These fits may not be
unique.  Much
of the H$_2$ emission from each star occurs in a narrow profile, with an intrinsic 
FWHM ranging from 17.5--28.5 \kms.  This narrow component of the H$_2$ emission
detected from TW Hya, DF Tau, and V836 Tau is centered at the radial
velocity of the star, but is shifted by $-12$, $-12$,
and $-27$ \kms\ in the spectra of  RU Lupi, T Tau, and DG Tau, respectively.  
The properties of this narrow component are presented in Table 11.
Additionally, a strong broad component, blueshifted from the primary narrow
component and extending to $-100$ \kms, is detected from RU Lupi and T Tau.
Weak blueshifted emission extending to $\sim-40$ \kms\ is also detected 
in both FUV spectra of DF Tau, which were obtained with a PA that differed 
by $180^\circ$.  Although the binarity of DF Tau may broaden the central profile, 
it cannot explain the broad blue wing. 

All of the H$_2$ emission lines from TW Hya, RU Lupi, DG Tau, and 
V836 Tau appear similar, regardless of the upper level.
Figure 11 shows that the asymmetric blueshifted wing from DF Tau and T Tau
is stronger from (1,4) and (1,7), relative to the flux in the narrow
component, than it is from (0,1), (0,2), and (2,12).
The  lines from (1,4) and (1,7)  from T Tau may also have additional
redshifted emission.  The \Lya\ emission profile that irradiates the the blueshifted gas
around DF Tau and T Tau, like that from HH objects, is narrower than the
\Lya\ emission profile that irradiates the bulk of the warm gas.

\subsection{Spatial Profile of the H$_2$ Emission}
Although the $0\farcs5\times0\farcs5$ and the
$0\farcs2\times0\farcs06$  apertures used in our observations are
narrow, the E140M echelle spectra yield some spatial information in the
cross-dispersion direction.  The pixels in the cross-dispersion direction
are 29 mas wide, so each pixel
corresponds to 4 AU at the 140 pc distance of the Taurus and Lupus
star-forming regions and 1.7 AU at the 57 pc distance of TW Hya.  These
two apertures are large enough that we are not limited by spatial resolution,
and small enough to minimize contamination by spatially extended gas
associated with the parent molecular cloud.

The coadded H$_2$ lines from DF Tau, RU Lupi, and T Tau are strong
enough to search for extended emission.
We do not analyze the spatial profile of H$_2$ emission from DG Tau or 
V836 Tau because their H$_2$ lines
are weak, and no atomic lines are strong enough to use for estimating the point-spread function.
\citet{Her02} found that the H$_2$ and \Lya\ emission from 
TW Hya were not significantly spatially extended relative to the hot lines such as \ion{C}{4}.

For each observation, we extract the spatially-resolved spectrum 
at the wavelengths of strong H$_2$ lines and various other strong lines
including the \ion{O}{1} 1305 \AA\ triplet, the 
\ion{C}{2} 1335 \AA\ doublet, the \ion{Si}{4} 1400 \AA\ doublet, and the
\ion{C}{4} 1550 \AA\ doublet. 
We subtract the spatial profile of the background, which is  measured from nearby spectral
regions with continuum emission but no detectable line emission.  
The central position of the emission in the cross-dispersion direction depends on the echelle order and the horizontal position of the wavelength
in that order and is automatically determined in the
{\it calSTIS} data reduction program.  When coadding the H$_2$ lines, we resample 
the emission profile in pixel space to account for sub-pixel differences.

The telescope thermal focus variations make the point-spread function of {\it HST}
uncertain (see \S 2).  The spatial profiles of the hot lines of \ion{C}{4} and \ion{Si}{4}
are most likely produced at or near the accretion shock \citep{Joh00,Cal04}
and should approximate the point-spread function.  The
strong lines produced in cooler 
gas, such as \ion{O}{1} and \ion{C}{2}, are also not spatially extended and can
 be used as a proxy for the point-spread function.  
The point-spread function of STIS can depend on wavelength, particularly
with the $0\farcs5\times0\farcs5$ aperture, so when possible we compare the
spatial profile of lines located near each other.  
This method is sensitive to 
emission extended beyond the hot emission lines, but is not sensitive to 
reflection of FUV emission from a disk surface or nebulosity.
Figure 12 shows the spatial distributions of various lines from DF Tau, T Tau, 
and RU Lupi.  We also construct space-velocity diagrams to analyze the spatial
distribution of co-added H$_2$ emission across the line profile.  

\subsubsection{DF Tau}
We concentrate on the observation of DF Tau
obtained with the $0\farcs5\times0\farcs5$ aperture,  because the S/N in
H$_2$ and other emission lines is higher than in the observation 
taken with the $0\farcs2\times0\farcs06$ aperture.  
All results are consistent with the
observations that used the small aperture. 

Figure 12 shows that the H$_2$ emission lines between 1270--1400 \AA\ (top left) and
1500--1650 \AA\  (bottom left) have a
similar profile to the \ion{C}{2} and
\ion{C}{4} emission lines, respectively.  The spatial profiles of this emission are well
characterized by the combination of a narrow and a broad Gaussian profile.
We measure an instrumental spatial resolution of $0\farcs10$ (3.6 pixels or 14.6 AU
at the distance of DF Tau) and $0\farcs14$ (4.7 pixels or 19.0 AU)
from the spatial profiles of \ion{C}{4} and \ion{C}{2}
emission, respectively.  
We identify weak asymmetric
emission in the wings of several lines, including \ion{C}{2} and \ion{C}{4}, offset
from the emission peak by about $0\farcs17$.
The spatial separation indicates that this emission is not directly
associated with the secondary star.   
The disk of DF Tau is observed close to edge-on.
Because dust strongly forward scatters FUV emission \citep{Dra03}, in
principle one side
of the disk could scatter such
extended emission.  This emission component may instead be an artifact from
an asymmetric point spread function. 

Figure 13 shows the space-velocity diagram for DF Tau for  the
observations obtained with the large aperture (left) and small
aperture (right).  The solid lines are
contours of 0.2, 0.5, and 0.8 times the peak flux for emission from
(0,1), (0,2), and (2,12).  The shaded
regions show the same contours for lines from (1,4) and (1,7).
The lines pumped near \Lya\ line center, from (1,4) and
(1,7), may be slightly more extended in the SW direction than the other
H$_2$ lines.

\subsubsection{RU Lupi}
Figure 12 compares the spatial extent of H$_2$ emission from RU Lupi between 
1270--1400 \AA\ (top) and 1500--1650 \AA\ (bottom) with that of \Lya,
\ion{Si}{4}, and \ion{C}{4}. 
The H$_2$ and \Lya\
emission is extended to the SW.  No other FUV emission lines, including
\Lya-pumped \ion{Fe}{2} emission, appear extended beyond a point source.
Figure 14 shows a space-velocity diagram of coadded H$_2$
emission from RU Lupi.  The shaded regions show contours of 0.2, 0.5, and
0.8 times the peak emission.  The solid lines show contours of 0.2, 0.5,
and 0.8 times the peak emission at each pixel in the spectra direction, and therefore
indicate the spatial distribution of emission across the line profile.  The
dashed lines indicate the point-spread function measured from \ion{Si}{4}
emission.  The H$_2$ emission at $\sim -15$ \kms\ is extended in the SW
direction but not in the NE direction.  About $70$\% of this emission at $\sim-15$ \kms\ 
is produced on-source.  We cannot discriminate between a
second point-source or continuous H$_2$ emission in the SW direction, but
some of the remaining 30\% of the emission must be extended by at least 70
mas from the star. 
The H$_2$ emission at $<-50$
\kms\ is not extended beyond the \ion{C}{4} emission (FWHM=0\farcs1).

Figure 15 shows the spatial extent of \Lya\ emission across the line profile
(solid line), smoothed by 75 \kms, compared to the spatial extent of
\ion{Si}{4} emission (dashed lines).
Like the H$_2$ emission at -15 \kms, the \Lya\ emission is spatially
extended SW of the star but not NE of the star.  
The spatial extent of the detected \Lya\ emission does not depend on the
velocity.  Even off source, the optical depth in the
\Lya\ line is large enough to produce emission at +800 \kms\ from line
center.

\subsubsection{T Tau}
Figure 12 shows that the spatial profile of coadded H$_2$ emission from
T Tau is slightly broader than the spatial 
profile of \ion{Si}{4} (top right) or
\ion{C}{4} emission (bottom right).
We coadd the H$_2$ lines from every upper level for this analysis to increase the S/N, 
even though the spectral profiles indicated that the lines
pumped near line center of \Lya\ have stronger blueshifted emission than
the other lines.  The space-velocity diagram
shown in Figure 16 indicates that the H$_2$ emission, particularly near
$v=0$ \kms, may be slightly more extended than the \ion{Si}{4} emission.  The emission at $v=-50$
\kms\ may be slightly less extended than \ion{Si}{4} emission.   We
caution, however, that the S/N in the space-velocity diagram is
low and that any conclusions drawn from the diagram may therefore be suspect.

\subsection{Wind Absorption of H$_2$ Lines}
Two strong H$_2$ lines, 0-4 R(0) and 0-4 R(1) at 1333.5 and 1333.8 \AA,
respectively, are located at $-240$ and $-165$ \kms\ from the
\ion{C}{2} 1334.5 \AA\ resonance line.  Those two H$_2$ lines may be
attenuated by the stellar wind if the optical depth in this
ground-state line is sufficiently large in the wind at those velocities.

Figure 17 shows the observed spectrum (shaded) and model H$_2$ emission
(dashed lines) for the 1333--1336 \AA\
spectral region for the six stars in our survey.  Table 12 describes the
wind properties and the attenuation of the 0-4 R(0) and 0-4 R(1) lines.  We
compare the maximum wind velocity ($v_\infty$) in the \ion{C}{2} 1334.5 \AA\ line with
that from the \ion{Mg}{2} 2796 \AA\ line.  We also compare the observed
flux in the two H$_2$ lines to the predicted flux, based on models
described in \S4.2.

\citet{Her02} found that the 0-4 R(1) 1333.8 \AA\ line from TW Hya is
weaker than expected from our models (see \S 4.2) because the wind
absorption in the \ion{C}{2} line extends to 
$\sim -180$ \kms\ in our line of sight.  Both the 0-4 R(0) and the
0-4 R(1) lines from RU Lupi are attenuated by the optically thick wind, which
extends to $\sim -240$ \kms.  On the other hand, the observed
flux in both lines from DF Tau is similar to the predicted flux, because
the optically thick \ion{C}{2} line in the wind of DF Tau only extends to $\sim -140$ \kms.
  Since the \ion{C}{2} wind absorption from T Tau extends to $\sim -180$
  \kms\ and the 
  H$_2$ emission is shifted by -12 \kms, the 
optical depth of the wind at the 0-4 R(1) 1333.8 \AA\ line
is small.  We
detect \ion{C}{2} emission between 1333.5--1333.8 \AA, so some of the
emission at 1333.8 \AA\ is likely from \ion{C}{2}.  Assuming the presence
of some \ion{C}{2} emission, we infer that the 0-4 R(1) line in the T Tau
spectrum is partially 
attenuated by the wind.  The fluxes in both H$_2$ lines from V836 Tau are
similar to the model flux, although we would not expect any optically thick
wind absorption  
at -165 \kms\ because the winds of WTTSs are not nearly as optically thick
at large velocities as winds from CTTSs. 
The 0-4 R(1) line at 1333.8 \AA\
appears weaker than expected from DG Tau, but the 0-5 R(1) and 0-5 R(2)
lines at 1394 \AA\ and the blend of 0-6 R(1) and 
0-6 R(2) at 1455 \AA\ are also much weaker than expected.   Therefore, we
cannot conclude that any H$_2$ absorption is attenuated by the
wind of DG Tau.

Several weaker H$_2$ lines also overlap with wind absorption features.
The 0-4 R(3) line at 1335.1 \AA\ line, located at $-130$ \kms\ from the
\ion{C}{2} 1335.7 \AA\ line, may be detected from DF Tau but is not detected
from TW Hya or RU Lupi because of wind absorption.  We did not include this
line from DF Tau in 
Table 5 because of the uncertain flux given the overlap with \ion{C}{2}
emission and low S/N.  The red side of the 0-4 R(3) line does not appear in the T
Tau spectrum, which strengthens our inference that the on-source H$_2$ emission from T
Tau is attenuated by the wind.  With better S/N than our data, several
other lines are potentially useful for measuring attenuation of H$_2$
emission by the wind, such as 3-2 P(16) at 
1305.663 \AA\ located at $-84$ \kms\ from the strong \ion{O}{1} 1306 \AA\
line.

These observations demonstrate that the H$_2$ emission detected from TW Hya, RU Lupi, 
and T Tau is absorbed by the the stellar wind.  Based on the similar patterns of H$_2$
emission, we infer that the H$_2$ emission would also be absorbed by the
winds of DF Tau, DG Tau, or V836 Tau if their winds were optically thick in \ion{C}{2} 
at larger velocities.
This attenuation suggests that the H$_2$ emission is produced
inside of any wind absorption, and as a result must be produced close
to the central star.

\subsection{Comparison to Previous Observations}
Table 13 compares the fluxes, FWHM, and velocities of H$_2$ emission from 
DF Tau, RU Lupi, T Tau, and DG Tau in our {\it HST}/STIS spectra and corresponding data obtained with {\it HST}/GHRS  of the same CTTSs, by \citet{Ard02a}.  
The {\it HST}/GHRS observations had $R=20,000$, compared to
$R=25,000-45,000$ with STIS, and used a
$2^{\prime\prime}\times2^{\prime\prime}$  
aperture, compared to $0\farcs2\times0\farcs06$ and
$0\farcs5\times0\farcs5$ with STIS.  The absolute velocities are 
accurate to $\sim20$ \kms\ for pre-COSTAR GHRS spectra and to $\sim3$ \kms\
for STIS spectra.
We compare H$_2$  
lines pumped on the red side and near \Lya\ line center.
Only a few line fluxes are co-added because the wavelength 
range of GHRS was limited to about 30 \AA\ per exposure.
We include the \ion{C}{4} flux as a rough proxy for the \Lya\ emission and
mass accretion rate. 

The centroid velocity of H$_2$ emission did not 
change significantly between the 
GHRS and STIS observations of RU Lupi, T Tau, and DF Tau.
The FWHM of the H$_2$ emission from RU Lupi and T Tau is broader in the 
GHRS observation than in the STIS observation, 
which could be produced either by a real velocity dispersion 
or by spatially extended emission in the dispersion direction.
The H$_2$ emission from the upper levels (1,4) and (1,7) from  DG Tau, 
RU Lupi, and T Tau was much stronger relative to the \ion{C}{4} emission 
during the GHRS observation than the STIS observation.  Because of the larger aperture used for the
GHRS observations than the STIS observations, 
we infer that a reservoir of warm molecular gas extending beyond the 
star was present for RU Lupi, T Tau, and DG Tau.

The H$_2$ emission from DG Tau was more blueshifted 
in our STIS observation than in the GHRS observation, 
but the two observations could have sampled
different gas.  
The GHRS observation of DG Tau exhibits 
strong \ion{C}{4} emission, which is not 
detected in our \STIS\ spectra even though we would expect such emission based 
on the flux ratio of H$_2$ to C IV.  The GHRS observations 
may have detected mostly extended emission in both H$_2$ and the other
lines, including \ion{C}{4}.  \citet{Gue05} detected extended X-ray
emission from jets emanating from DG Tau.  Hot gas traced by 
emission in \ion{C}{4}
\citep{Ray97} and possibly \ion{O}{6} \citep{Her05} has previously been
detected in outflows from CTTSs.
However,
\citet{Wal03} and \citet{Sau03} did not detect any extended \ion{C}{4} emission
from T Tau, despite the presence of extended X-ray emission \citep{Gue04}.
\citet{Tak04} found near-IR H$_2$ emission in several lines from DG Tau, 
including the 1-0 S(1) transition, which is blueshifted by
$\sim 15$ \kms\ from the system velocity and offset from the source by
$0\farcs2$.  They used relative H$_2$ line fluxes to estimate a
temperature of 2000 K, which is hot enough to produce
\Lya-pumped H$_2$ emission in the FUV \citep[e.g.][]{Bla87,Her04}.  
\citet{Ric02}
estimated an upper flux limit of $3\times10^{-14}$ \erg\ in the S(2) line
at 12 $\mu$m from DG Tau.  

\citet{Naj03} detected fundamental CO emission from V836 Tau, but not from
V410 Tau.  \citet{Duv00} also detected weak CO $J=2-1$ emission from V836
Tau.   \citet{Bar03} placed a flux upper limit of $9.4\times10^{-16}$ \erg\
for
the 1-0 S(1) line from V836 Tau. Our detection of H$_2$ emission from V836
Tau but not from V410 Tau is therefore consistent with previous studies
that suggested that V836 Tau retains gas in its disk but that V410 Tau retains neither a
gas nor a dust disk.

\section{Discussion}

\citet{Her02} found that the fluorescent H$_2$ emission from TW Hya is
symmetric about the radial velocity of the star and is not spatially
extended.  They also found that the H$_2$ emission from TW Hya is
attenuated by the wind,  
which constrains the origin of the emission to be close to the star. The
only molecular  
gas known to be associated with TW Hya resides in its circumstellar disk.
Therefore, the warm
disk surface is a likely source for the H$_2$ fluorescence.
However, the diverse spectral and spatial profiles of H$_2$ emission
from other TTSs indicate that the source of H$_2$ emission 
depends on the target. Figure 2 shows no
H$_2$ absorption against the observed \Lya\ emission from TW Hya, RU Lupi, or DF Tau.  
The \Lya\ emission from Mira B, which excites a similar pattern of H$_2$ fluoresence as seen here, shows several H$_2$ absorption because the warm H$_2$ is in our line of sight to the \Lya\ emission source \citep{Woo02}.  Thus, the warm H$_2$ is not in our line of sight to TW Hya, RU
Lupi, or DF Tau.
We note that TW Hya, T Tau, and RU Lupi have disks that are
most likely  viewed face-on, whereas DF Tau and DG Tau have disks that are
viewed edge-on.  Table 14 summarizes stellar properties and the properties
of the H$_2$ fluorescence for each source.

The H$_2$ emission from T Tau, RU Lupi, and DG Tau is blueshifted by 
10--30 \kms, which is similar to the blueshifts of H$_2$ emission from 
HH objects \citep{Schw03} and indicates an outflow origin.
The velocity of FUV
H$_2$ emission from DG Tau is about 10 \kms\ larger than that of the 
IR H$_2$ emission, which is offset by $0\farcs2$ from the star and
clearly assosciated with an outflow
\citep{Tak04}. 
The
blueshifted H$_2$ emission from T Tau and RU Lupi extends to $-100$ \kms\ and is not
spatially extended.  Detecting H$_2$ at such high velocities is surprising
because a strong shock should destroy the H$_2$, given its dissociation
energy of 4.5 eV.   
When produced in an outflow, the H$_2$ emission appears blueshifted 
regardless of whether the disk is observed edge-on or face-on.  The
asymmetic blueshifted emission, however, extends to a much larger velocity
for the face-on CTTSs T Tau and RU Lupi than for the edge-on CTTS DF Tau.
The absorption of H$_2$ emission by \ion{C}{2} in the wind restricts the
H$_2$ to be produced interior to optically thick wind absorption.
Therefore, the H$_2$ emission from these sources is likely produced at or
near the base of the outflow.

Figure 18 shows the space-velocity diagram of H$\beta$ emission observed in our 
STIS G430L observation of RU Lupi.  The blueshifted  H$\beta$ emission is
spatially extended 
symmetrically about the star, as we would expect for emission from a stellar
wind, a disk wind from a face-on star, or possibly from the disk itself. 
The SW extent of the H$_2$ and \Lya\ emission suggests that they are instead
produced by a jet, possibly sweeping up nearby molecular material.  \citet{Tak01} used spectro-astrometry of RU Lupi to find that 
blueshifted H$\alpha$ emission is displaced by 20--30 mas to the SW of the star, and blueshifted
[\ion{O}{1}] and [\ion{S}{2}] emission is displaced by 30--300 mas to the
SW of the star.
\citet{Tak01}, however, also find redshifted H$\alpha$ emission displaced by 30 \kms\
to the NE.  Based on this detection, they infer that the edge-on disk of RU
Lupi may have a central hole of 3--4 AU.  This observational
result is difficult 
to reconcile with the absence of either H$_2$ or \Lya\ emission extended NE
from the star and the absence of any extended redshited H$\beta$ emission.

\citet{Gra05} found that the FUV emission from
several CTTSs is extended by up to an arcsecond.
This spatially extended emission may be produced either by extended H$_2$
gas, forward
scattering by dust in a nearby nebulosity,
 or a jet interacting with nebulosity.  
In cases where only the H$_2$ emission from the CTTSs is extended, it is
most likely related to the stellar outflows rather than the disk. 
The blueshifted H$_2$ emission may be produced where nearby nebulosity is 
shocked by outflows or in the dense outflows 
from CTTSs \citep[e.g.,][]{Gom01}.

In their long-slit FUV spectra, \citet{Wal03} and \citet{Sau03}  
found that fluorescent H$_2$ emission is extended by at least
8$^{\prime\prime}$ from T Tau.  This emission is most likely produced by
the stellar outflows where they shock the surrounding molecular cloud.   
In the off-source spectrum, the only detected lines are from (\Vup,\Jup)=(1,4)
and (1,7), which  
are pumped near \Lya\ line center, as is also the case for HH43 and HH47
\citep{Sch83,Cur95}. 
The presence of emission in these two progressions and
absence of emission in any other progression imply that the \Lya\
emission is relatively 
narrow when produced by outflows that shock the 
molecular gas.  However, when $\log N($\ion{H}{1}$)>14.5$, the \ion{H}{1}
absorption at $+14$ \kms\ is optically thick.  The \Lya\ emission that
excites (1,7) must therefore be produced {\it in situ}.
On the other hand, \citet{Wal03} found in their long-slit spectra of T Tau
that the on-source  
H$_2$ emission has a much richer spectrum, which we confirm with our
echelle spectrum.   
The profile of the \Lya\ emission line that excites the on-source H$_2$ emission from T Tau 
must be broad and may be produced by accreting gas.

Like the extended emission from T Tau, we find 
excess blueshifted emission only 
in the lines from (1,4) and (1,7) in the on-source spectra of DF Tau and T Tau.
 The \Lya\ emission that irradiates
this gas must be narrower than the \Lya\ emission that irradiates the
bulk of the on-source molecular gas and may be produced in the accretion
shock.  The narrow \Lya\ emission associated with the outflow must also be
produced close to the warm, blueshifted H$_2$ emission.

We do not detect any differences in the H$_2$ emission profiles from the various
upper levels in the RU Lupi spectrum, even though the spatial extent of the
H$_2$ and \Lya\ 
emission implies that they are produced in 
the outflow.  The outflow of RU Lupi is
sufficiently optically thick that 
extended \Lya\ emission is seen at $+800$ \kms. 
In contrast, the absence of spatially extended \Lya\ emission from TW Hya
and DF Tau implies 
that the \Lya\ emission from those two sources is most likely produced by
the accreting gas.  Because RU Lupi has a high mass 
accretion rate, the accretion flow may be optically thick to \Lya\
emission, preventing us from detecting the \Lya\ emission produced by the accreting gas.  
\citet{Sta04} speculate that X-ray emission from strongly accreting CTTSs
may be similarly attenuated by accreting gas. 
 The strong outflow that produces \Lya\ emission from RU Lupi may also
 produce the blueshifted \ion{O}{6} 1035 \AA\ and 
\ion{C}{3} 977 \AA\ emission detected in {\it FUSE} spectra, and could
contribute to the complicated emission profiles of \ion{Si}{4} and
\ion{C}{4} \citep{Her05}.

The bulk of the H$_2$ emission from DF Tau and V836 Tau is consistent with
a disk origin, because the H$_2$ emission has the same radial velocity as
the star and the bulk of the emission from DF Tau is not spatially extended.  Our observations are
unable to determine the spatial extent of H$_2$ emission from V836 Tau.
The weak blueshifted H$_2$ emission from DF Tau is most likely associated
with an outflow.   The total H$_2$ flux from TW Hya, DF Tau, and V836 Tau is not 
an appropriate indicator of the total amount of H$_2$ present in
circumstellar material, since the FUV H$_2$ emission is produced only in
warm ($2000-3000$ K) molecular gas that is irradiated by a strong 
FUV emission source and that has a large filling factor around that source.

Irradiated by strong \Lya\ emission, the hot disk surface of CTTSs can
produce the observed H$_2$ fluorescence.
Therefore, it is surprising that very little if any H$_2$ emission in the
spectra of T 
Tau, RU Lupi, and DG Tau is produced at the stellar radial velocity, which
suggests that 
fluorescent H$_2$ emission is not produced at the surface of their disks.
T Tau, RU Lupi, DG Tau, and DF Tau 
all have large mass accretion rates, while TW Hya has a small mass accretion rate.  
V836 Tau retains a disk and may be weakly accreting.
\citet{Ard02a} found that the H$_2$ emission is not blueshifted from BP Tau
and RY Tau,  
both of which have moderate mass accretion rates, or from RW Aur, which has a high mass 
accretion rate.  The H$_2$ lines from the strongly accreting star DR Tau are 
blueshifted by 10 \kms.  Moreover, the velocity shift of the H$_2$ emission does not
appear to be correlated with dust settling or disk clearing, which is
identified by the absence of excess NIR emission.

Mass loss rates from CTTSs scale with mass accretion rates \citep{Har95}.
The H$_2$ emission from the strong
accretors tend to be blueshifted, and may be produced
by outflows sweeping up molecular gas that either surrounds the star or is
located at the disk surface, or at the base of the wind.  Most of the \Lya\ emission is likely
produced by the accreting gas, but it may also be produced or scattered by
outflows.
The absence of H$_2$ emission from the disks of most of these strong accretors could result 
from the accretion flow being optically thick to the \Lya\ emission that is 
produced by the accreting gas. 

\citet{Bla87} calculated the FUV and IR H$_2$ emission produced by \Lya\
pumping of H$_2$.  The FUV pumping of H$_2$ can change the level
populations and, as a result, modulate the emission in IR H$_2$ lines.
The models calculated by \citet{Bla87} included a relatively narrow \Lya\ line because the {\it IUE} data
from T Tau \citep{Bro81} and HH objects \citep{Sch83} showed
emission only in the H$_2$ lines pumped near line center.  Although their
generic description of the H$_2$ fluorescence from CTTSs is accurate, their specific results may not
necessarily apply to the rich
on-source H$_2$ emission spectra described here.  \citet{Nom05} revisited the
relationship between FUV pumping and IR H$_2$ emission with a more
realistic \Lya\ profile from TW Hya and a disk geometry to explain the
emission in the strongest progressions.  Their models predict
H$_2$ excitation temperatures that are too low to explain the emission in
several other progressions, but may be sufficient with additional heating by X-rays.  These investigations also demonstrate that
emission in IR lines from warm H$_2$ gas depend on the FUV emission.

\section{CONCLUSIONS}
We have analyzed H$_2$ fluorescence in high-resolution \HST/STIS echelle spectra
of the CTTSs TW Hya, RU Lupi, DF Tau, T Tau, DG Tau and the weakly
accreting WTTS V836 Tau
with the following results: 

1.  Between 41--209 H$_2$ lines are detected in each of these far-UV spectra.  The
H$_2$ emission 
is much brighter from TW Hya, RU Lupi, and DF Tau than from T Tau, DG Tau,
and V836 Tau. 
The strength of H$_2$ emission depends on the amount of warm (2000--3000 K)
H$_2$ gas in the environments around the 
various stars, the strength of \Lya\ emission, 
and the solid angle filling factor of H$_2$ around the \Lya\ emission. 
This emission does not trace the bulk
of the gas in the disk, which is cold.

2.  The H$_2$ lines are pumped from many
different levels, with the strongest lines typically pumped by five transitions located 
between 0 to 550 \kms\ from \Lya\ line center.  Several H$_2$ lines from all sources
except T Tau are pumped by blueshifted \Lya\ emission, although those lines 
are weak in the spectra of DF Tau, RU Lupi, and V836 Tau.  The H$_2$ lines from
DF Tau and T Tau suggest that the \Lya\ emission irradiating the warm
molecular gas around those stars is redshifted.

3.  In the spectra of TW Hya and DF Tau, we find H$_2$ emission pumped by \ion{C}{4} from highly 
excited lower levels.  This H$_2$
emission is consistent with $\sim3$\% of the warm gas 
having an excitation temperature of $\sim 2.5\times10^{4}$ K.  The highly-excited gas could 
be related to the strong FUV continuum detected from TW Hya and DF Tau.  RU Lupi 
shows neither a strong FUV continuum nor these highly excited lines.
Surprisingly, many highly-excited H$_2$ lines 
that could be pumped by \Lya\ are not detected.

4.  H$_2$ emission is detected from the WTTS V836 Tau, which retains a
dust disk and is weakly accreting, but not from the WTTS V410 Tau, which no longer has a disk.

5.  \citet{Wal03} found that \Lya\ emission from T Tau is present in two G140L 
long-slit spectra but is absent in another long-slit spectrum and in the E140M 
echelle spectrum.  We use the variability in H$_2$ lines from different upper 
levels to demonstrate that the width of the intrinsic \Lya\ profile changes, 
while the extinction and \ion{H}{1} column density remain constant.  We
also find that the presence of H$_2$ emission pumped by the red wing of
\Lya\ is correlated with \Lya\ emission at those wavelengths, even though
our viewing angle is different than that seen by the fluorescing H$_2$ gas.
We therefore conclude that the presence or absence of emission in the red
wing of the \Lya\ emission profile is
isotropic.

6.   With the possible exception of DG Tau,
    the H$_2$ emission studied here must be produced close to the star
    because the bulk of the emission is not extended beyond a point source.
    The absorption of H$_2$ emission by \ion{C}{2} in the wind requires
    that the H$_2$ emission is produced interior to the optically thick
    wind absorption.  In cases where the \Lya\ is observed, the H$_2$ gas is
    not located in our line of sight to the \Lya\ emission region.  The on-source H$_2$ fluorescence
    is excited by broad \Lya\ emission, in contrast to the spatially extended H$_2$
    fluorescence detected from HH objects and the molecular complexes
    surrounding T Tau.

7.  The H$_2$ lines show a diverse range of spatial and spectral profiles.
The H$_2$ emission from RU Lupi, T Tau, and DG Tau is blueshifted,
suggesting an outflow origin, while the emission 
from TW Hya, DF Tau and V836 Tau is centered at the radial velocity of
the star, suggesting a disk origin.  The H$_2$ lines from DF Tau, RU Lupi, and T
Tau include a weak blueshifted component.

8.  In the spectra of T Tau and DF Tau, the H$_2$ lines from the
(\Vup,\Jup)=(1,4) and (1,7) upper levels,  
which are pumped near \Lya\ line center, have stronger excess blueshifted
emission, relative to the flux in the narrow component, than the other H$_2$ lines pumped 
further from line center.  The blueshifted component of these lines
is likely produced in the outflow and may be pumped by \Lya\
emission produced {\it in situ} rather than in the accreting gas.

9.  The H$_2$ and \Lya\ emission from RU Lupi are both extended to the SW
relative to the 
emission in lines of \ion{C}{4} and \ion{Si}{4}, which are most likely
produced by accreting gas.  The H$_2$ and \Lya\ emission is presumably related to the
blueshifted H$\alpha$, [\ion{O}{1}], and [\ion{S}{2}] emission also detected to
the SW of the star by \citet{Tak01}.
We speculate that the observed \Lya\ emission from RU Lupi, which pumps
the H$_2$, may arise entirely in the outflow.   The \Lya\ emission produced
by the accreting gas could be absorbed by the accretion column.

10.  Comparison of our STIS echelle spectra with the GHRS observations suggests 
that H$_2$ emission from T Tau, RU Lupi, and DG Tau is also extended, particularly 
from the levels (1,4) and (1,7), which are pumped near \Lya\ line center.
Extended H$_2$ is likely produced by outflows that shock the  surrounding
molecular material. 

11.  We do not find any stellar property that reliably predicts whether the H$_2$
     emission will be produced in the warm disk surface or the outflow.
     The presence or absence of blushifted 
     H$_2$ emission does not appear to depend on disk evolution or
     environment.  The absence of H$_2$ emission from a disk, however, tends to occur
     for the stars with higher mass accretion rates.

\section{Acknowledgements}
This research was funded in part by STScI programs GTO-7718, GTO-8041, and
GO-8157 to the University of Colorado and to SUNY Stony Brook, 
and by the Swedish National Space Board. This paper is based on
observations made by the NASA/ESA Hubble Space Telescope, obtained at the
Space Telescope Science Institute, which is operated by the Association of
Universities for Research in Astronomy, Inc., under NASA contract
NAS5-26555.  This research was also funded in part by the STScI program
AR-9930 to the University of Colorado and by NASA program S-56500-D to the
University of Colorado.

We thank the anonymous referee for valuable suggestions.
We also thank Ilya Ilyin, who obtained our  {\it NOT} observations  and reduced
the spectrograms.  GJH thanks Carol Grady for valuable discussion
concerning the spatial distribution of emission from several other CTTSs,
and Gail Schaefer for valuable discussion regarding the position of DF Tau
in the aperture.



\clearpage

\begin{figure}
\label{calflux_ttau.eps}
\plotone{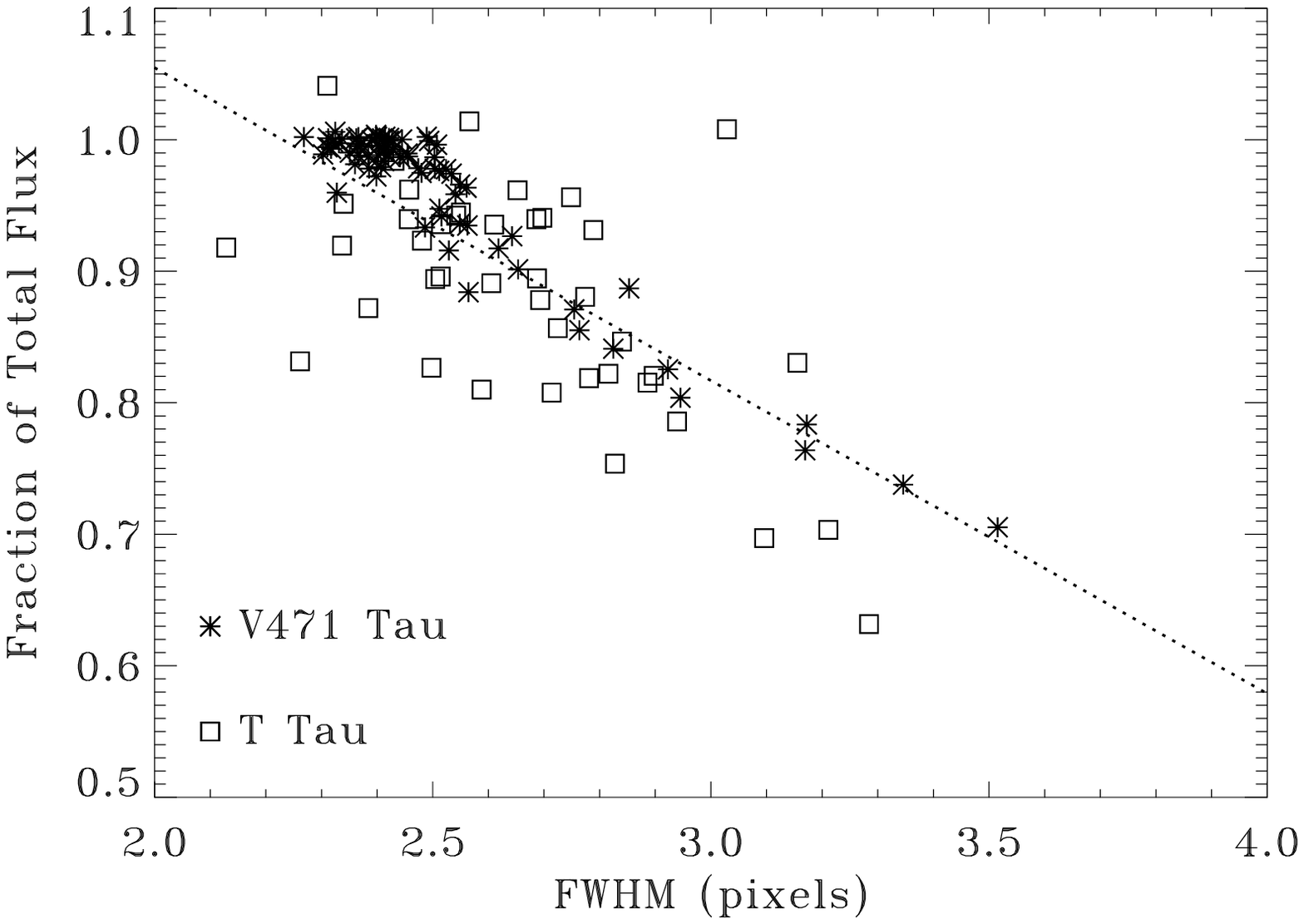}
\caption{The dependence of count rate on the telescope focus.  Changes in
  the telescope focus due to thermal variations alter 
the count rate in our
observations of T Tau because the point-spread function may be wider than the
aperture.  We calibrate the observed flux by comparing the measured
point-spread function and the flux in 300s time intervals.
The point-spread function is the average of the point-spread
functions measured in four spectral regions with strong emission lines in
the T Tau spectrum (\ion{O}{1} 1305 triplet, \ion{C}{2} 1335 \AA\ doublet,
\ion{Si}{4} 1400 \AA\ doublet, and \ion{C}{4} 1549 \AA\ doublet).  The FUV
spectrum of V471 Tau is dominated by continuum emission from the white
dwarf.  We find the best-fit slope from the V471 Tau data and scale the T
  Tau data to calculate the total flux.} 
\end{figure}

\begin{figure}
\epsscale{0.7}
\plotone{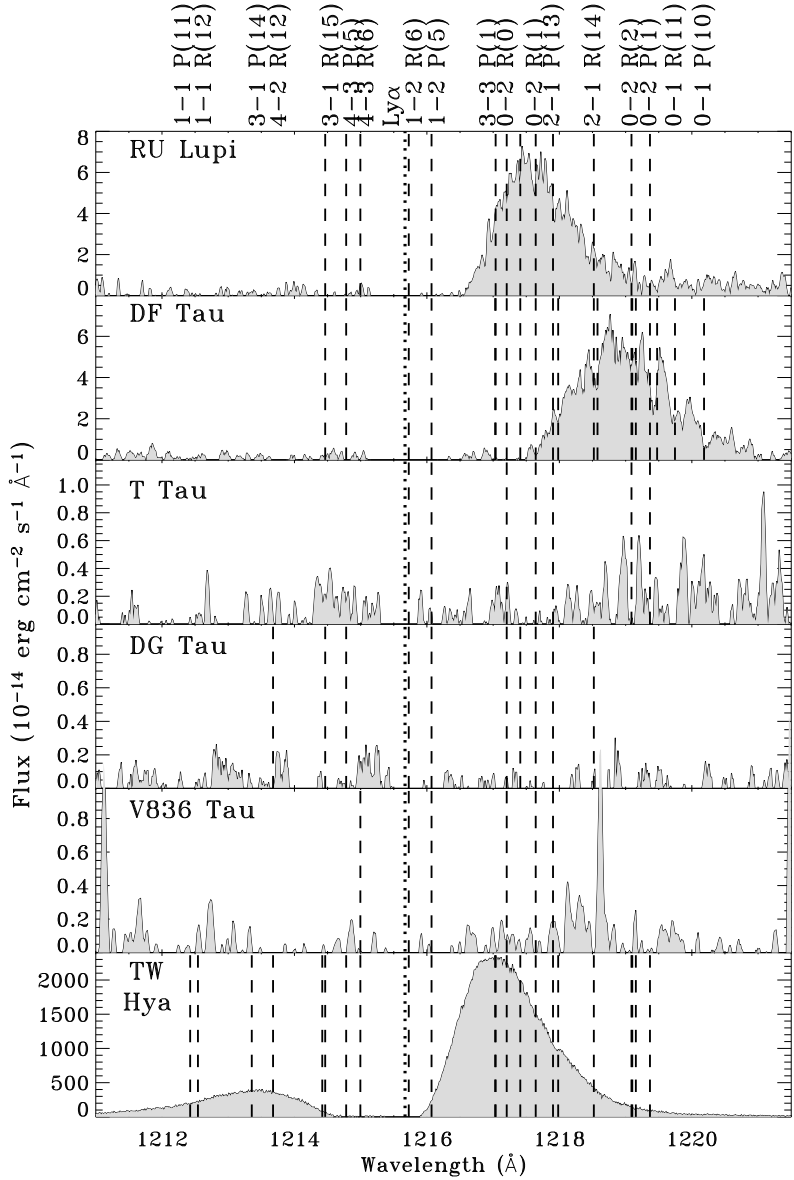}
\caption{Observed \Lya\ emission from TTSs (shaded).  The detected \Lya\ emission from TW Hya is strong 
because the \ion{H}{1} column density in our line of sight is small.  DF
Tau and RU Lupi also show some \Lya\ emission, while the \Lya\ emission
from T Tau, DG Tau, and V836 Tau is completely attenuated in our line of sight by
\ion{H}{1} in the wind and interstellar medium.  The vertical dashed lines, most of
which are labeled at the top, indicate
the transitions that excite H$_2$ emission for each star.  The broad \Lya\
emission from DF Tau pumps H$_2$ emission primarily in its red wing.  Except for TW Hya, the stars show few or no H$_2$ lines pumped by emission shortward of \Lya\ line center.}  
\end{figure}

\begin{figure}
\plotone{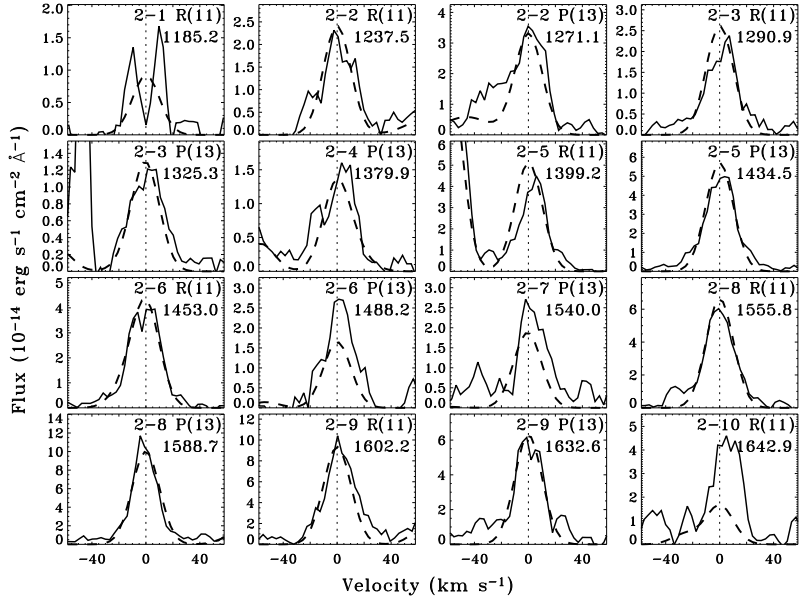}
\caption{16 H$_2$ lines from the upper level (\Vup,\Jup)=(2,12) in the STIS observation of DF
Tau obtained with the $0\farcs5\times0\farcs5$ aperture.  The predicted
relative fluxes (dashed lines) described in \S 4.2 are scaled to the
observed fluxes and fit the observed emission (solid lines) well.  
Several lines, such as the 2-1 R(11) line at 1185 \AA, are weak and only identified
as H$_2$ because the models predict the presence of emission at these
wavelengths.  The 2-10 R(11) line at 1642.9 \AA\ may be blended with an
\ion{Fe}{2} line.}
\end{figure}

\begin{figure}
\epsscale{0.8}
\plotone{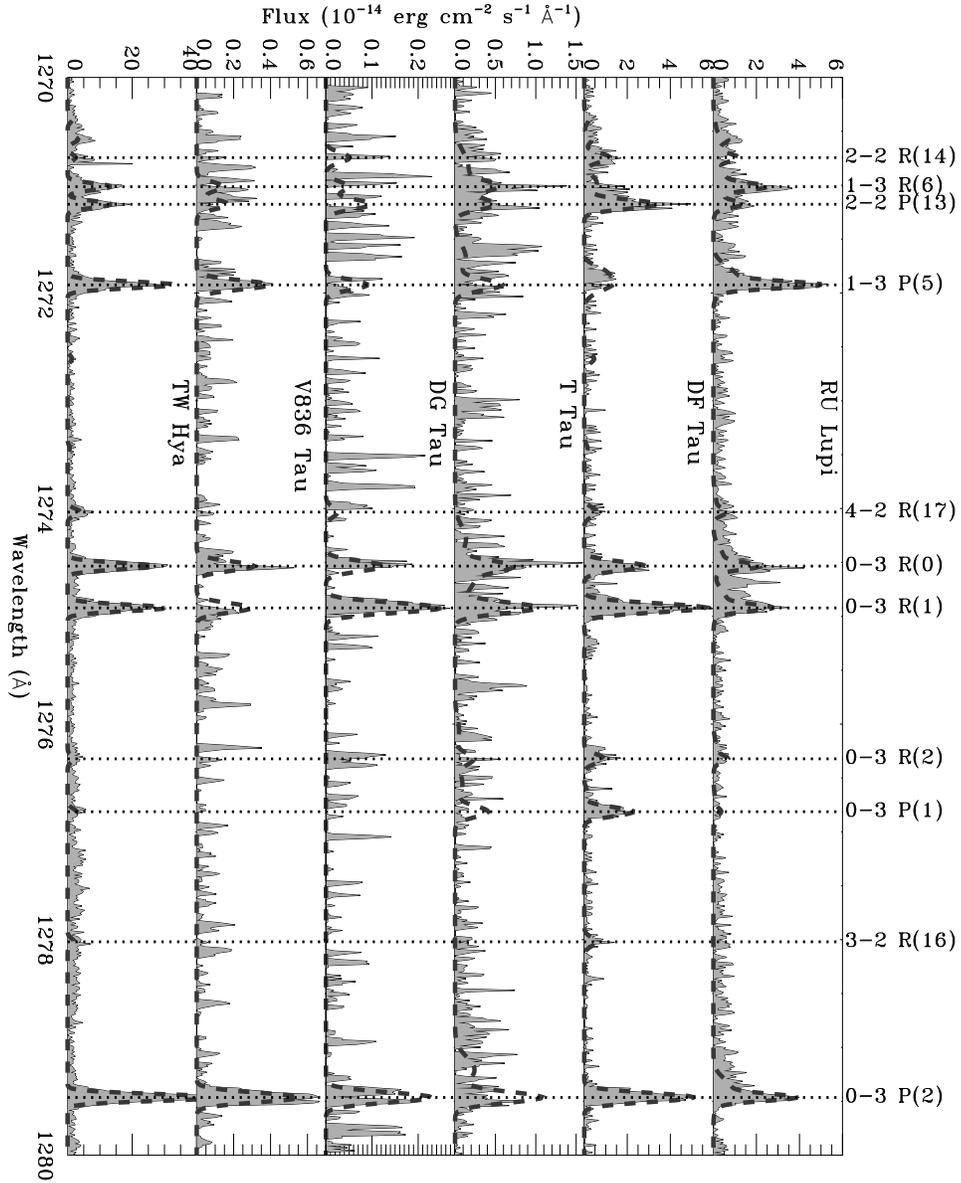}
\caption{The spectral region from 1270--1276 \AA\ for six stars.  We
  display the observation of DF Tau obtained with the
  $0\farcs5\times0\farcs5$ aperture.  The observed emission (shaded) is
  shifted in wavelength so that the H$_2$ lines occur at the theoretical
  wavelengths 
calculated by \citet{Abg93}.  The model H$_2$ spectrum (dashed line),
described in \S 4.2, well fits the observed H$_2$ lines, several of which
are identified.}
\end{figure}

\begin{figure}
\epsscale{0.8}
\plotone{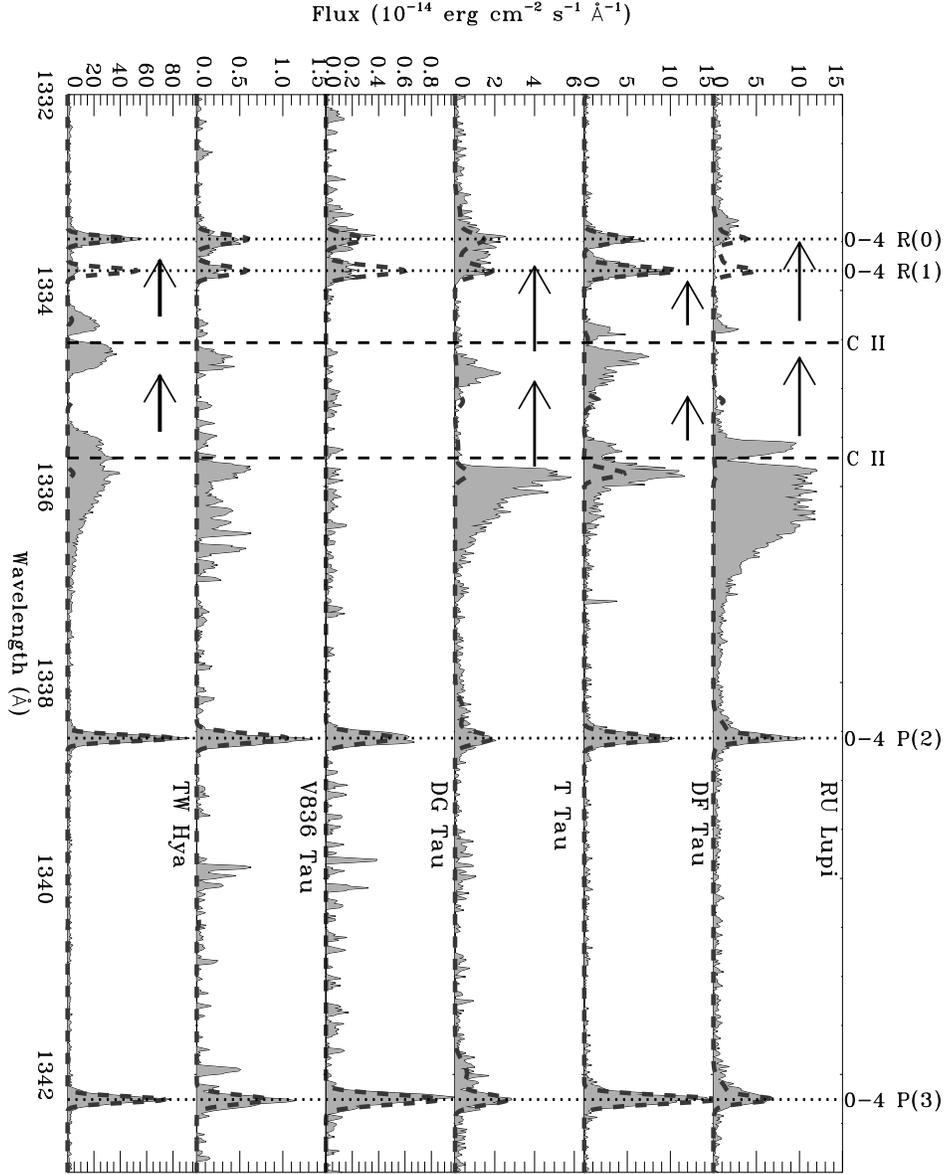}
\caption{Same as Figure 4 for the spectral region 1332--1343 \AA.  This
region is dominated by the \ion{C}{2} doublet, as seen in strong redshifted
emission and blueshifted absorption.  The wavelength extent of wind absorption in both
the \ion{C}{2} 1334.5 and the \ion{C}{2} 1335.7 \AA\ lines is shown by the
arrows.  The narrow absorption feature near line center of \ion{C}{2} is
produced by the interstellar medium.}
\end{figure}

\begin{figure}
\epsscale{0.8}
\plotone{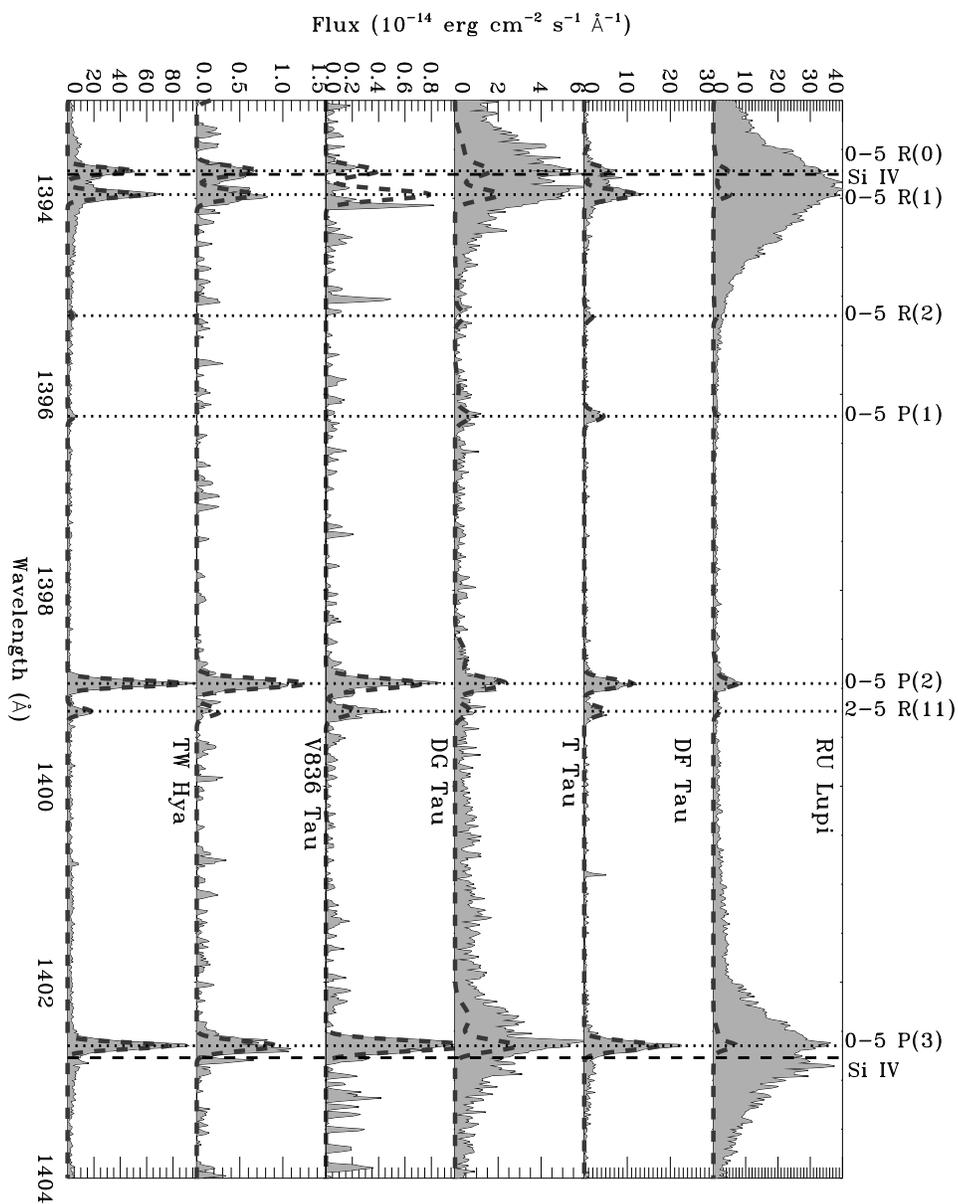}
\caption{Same as Figure 4 for the spectral region 1393--1404 \AA.  The
\ion{Si}{4} resonance doublet can be quite strong and mask several H$_2$
emission lines, as is seen in the RU Lupi and T Tau spectra. In the spectra of DF Tau and TW
Hya, \ion{Si}{4} emission is weak and the region is instead dominated by
strong H$_2$ lines.}
\end{figure}

\begin{figure}
\epsscale{0.8}
\plotone{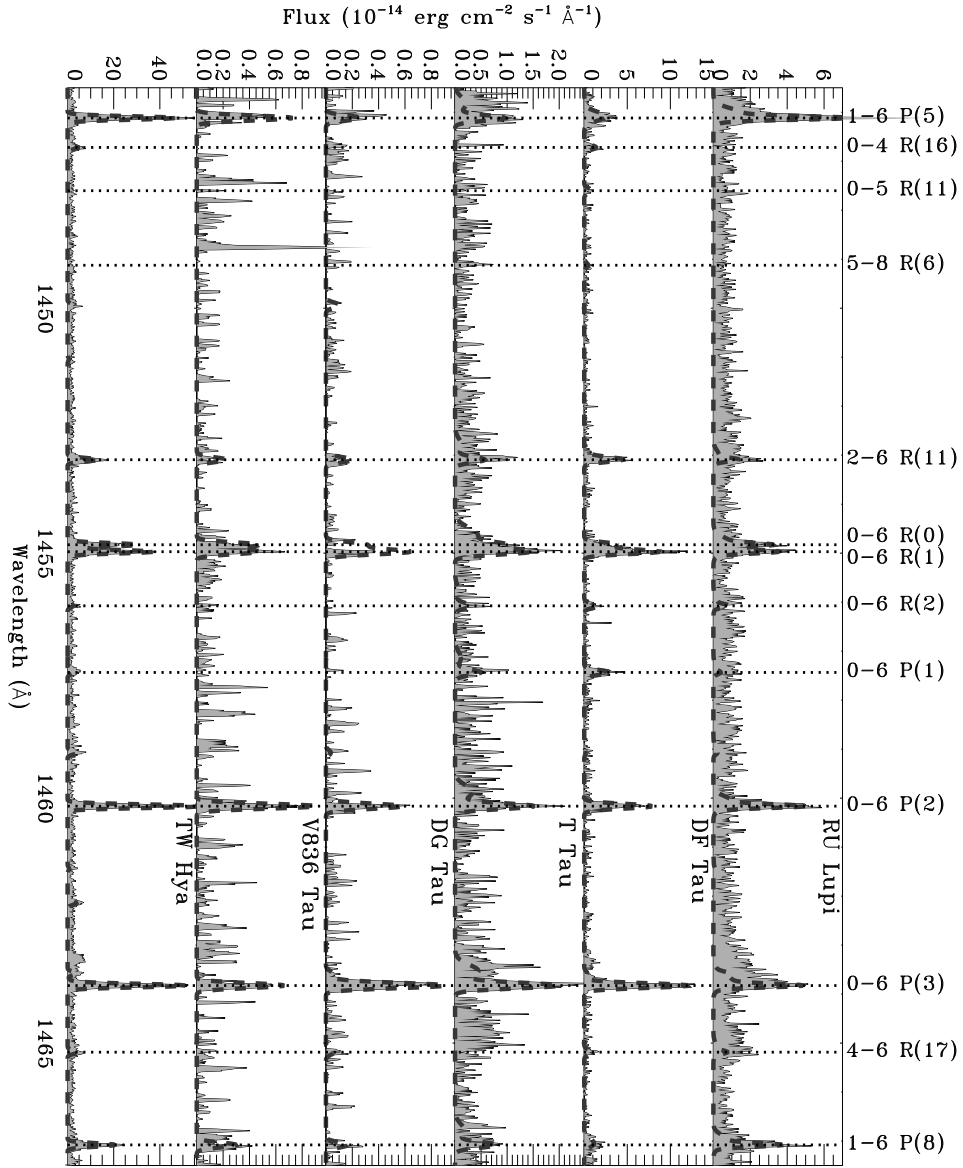}
\caption{Same as Figure 4 for the spectral region 1453--1467 \AA.}
\end{figure}

\begin{figure}
\epsscale{0.7}
\plotone{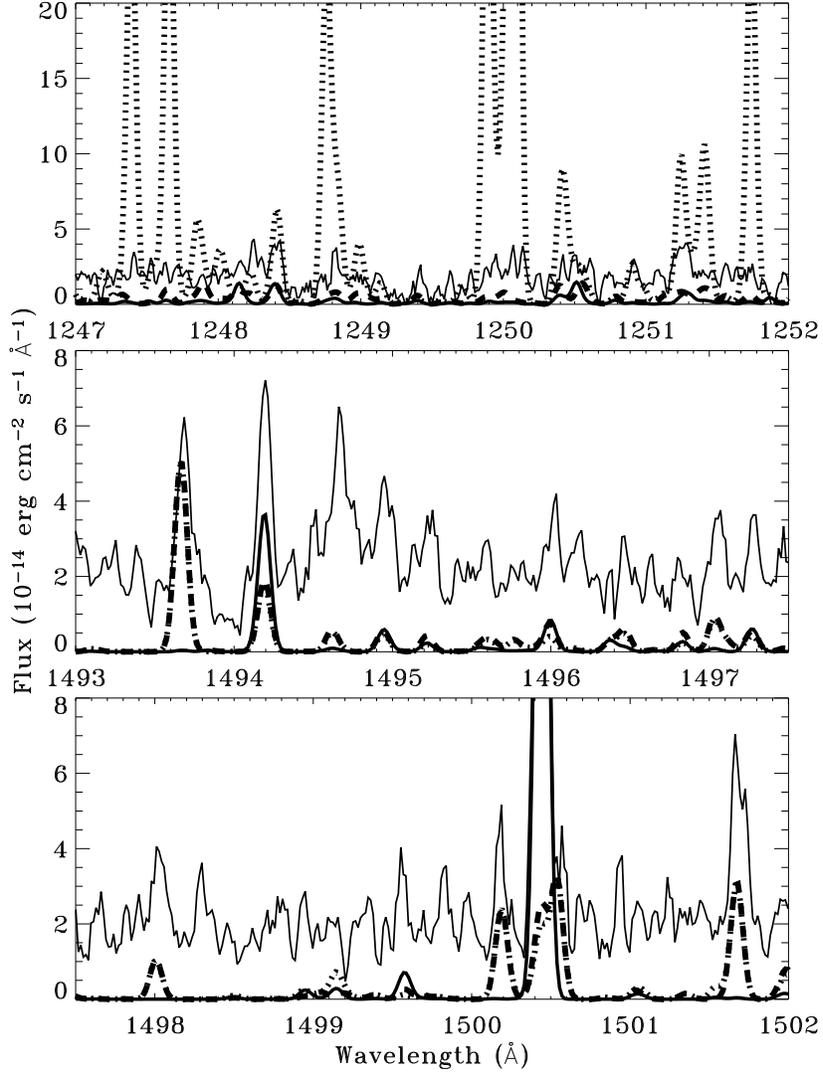}
\caption{Synthetic H$_2$ emission spectra, estimated by calculating FUV pumping, compared to the observed spectrum of TW Hya (thin solid line).
The synthetic spectra are calculated by
  simulating H$_2$ fluorescence in a plane-parallel slab of H$_2$ with
  $T=2500$ K and $\log N$(H$_2$)=18.5 (thick solid line) and two models an
  additional layer of highly excited H$_2$ emission with  $T=2.5\times10^4$
  K and $\log N$(H$_2$)=17 (dashed and dotted lines).  All of the models include pumping throughout the FUV, although one model (dashed lines) does not include any pumping by \Lya.
Lines at
  1493.7, 1498.0, 1500.2, and 1501.7 require highly excited H$_2$ gas.  However, if
  \Lya\ irradiates this highly excited gas, we would expect to see many
  lines that are not present in the data (dotted lines in top panel).}
\end{figure}

\begin{figure}
\epsscale{0.7}
\plotone{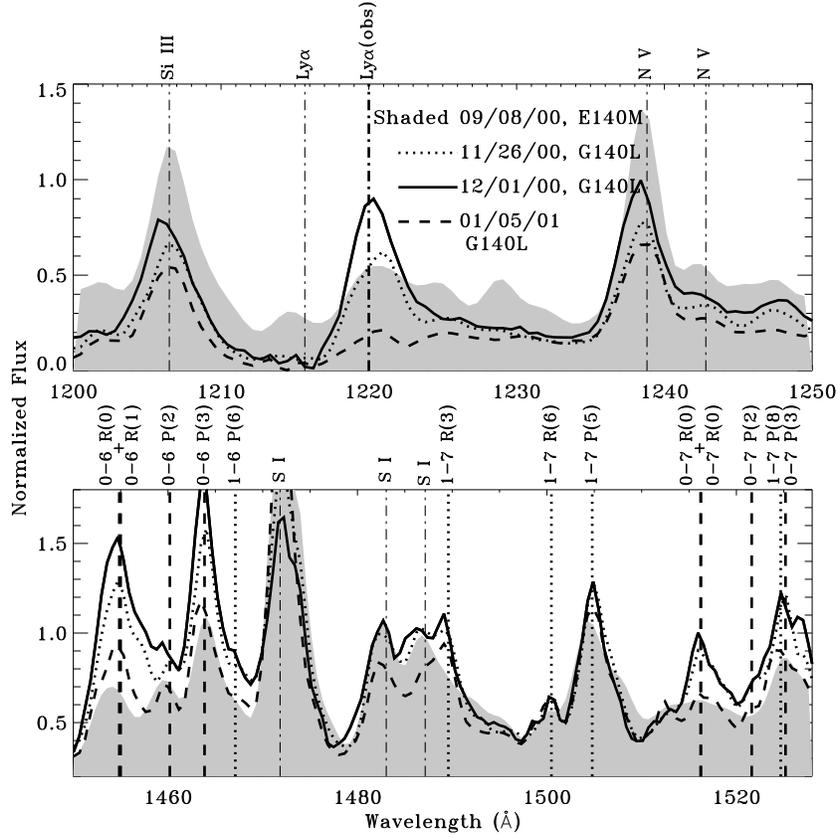}
\caption{\Lya\ and H$_2$ emission in four spectra of T Tau, scaled by the flux 
between 1495--1510 \AA.  We previously obtained three long-slit G140L spectra of 
T Tau, with dates labeled \citep{Wal03}.  We convolve the E140M echelle 
spectrum (shaded) to the resolution of G140L ($R=1000$).  The \Lya\ emission is 
strong in the observations obtained on 11/26/00 and 12/01/00 but is not present 
in the echelle spectrum (the apparant bump at 1220 \AA\ is noise, see Fig. 2) or 
the spectrum obtained on 01/05/01.  As can be seen at 1455--1465 \AA\ 
and 1515--1525 \AA, when \Lya\ emission is not present the H$_2$ emission from (\Vup,\Jup)=(0,1) and (0,2), marked by dashed vertical lines are much weaker relative to the H$_2$ emission 
from (1,4) and (1,7), marked by the dotted vertical lines.
Because lines from (0,1) and (0,2) are pumped by the red wing of \Lya, we infer that the 
intrinsic \Lya\ profile is narrower during those observations.}
\end{figure}

\begin{figure}
\plotone{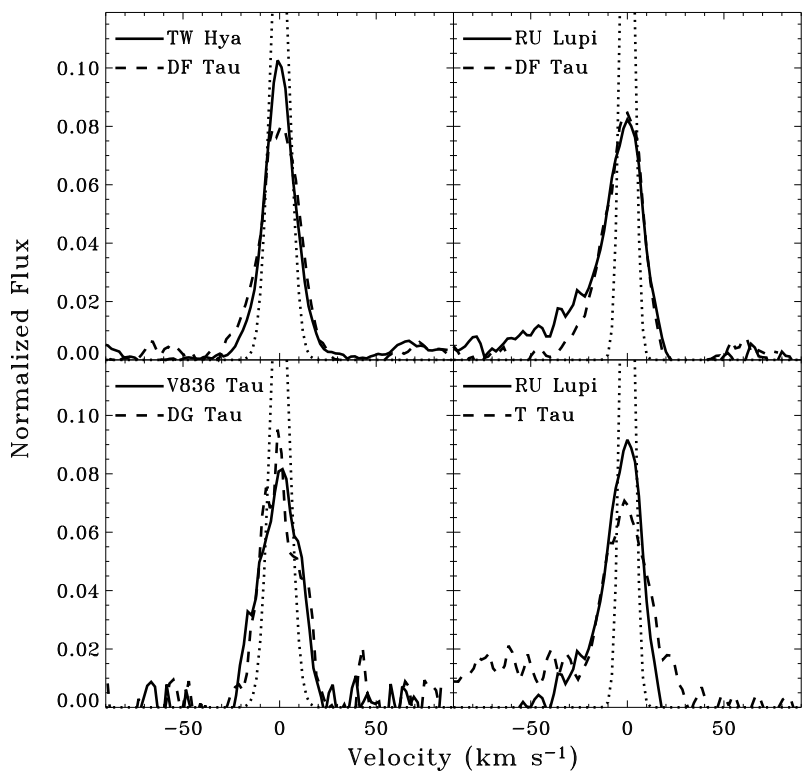}
\caption{Spectral profiles of H$_2$ lines from upper levels (\Vup,\Jup)=(0,1), 
(0,2), and (2,12), shifted in velocity so that line center is
at $v=0$.  The H$_2$ lines from RU Lupi, DF Tau, and T Tau all show a
significant blueshifted component.  Dotted lines indicate the instrumental line-spread 
function.  The blueshift from DF Tau is evident in both observations, which had PA that 
differed by 180$^\circ$, so the blueshift is real and not a spatial displacement of the secondary star in the dispersion direction.}
\end{figure}

\begin{figure}
\plotone{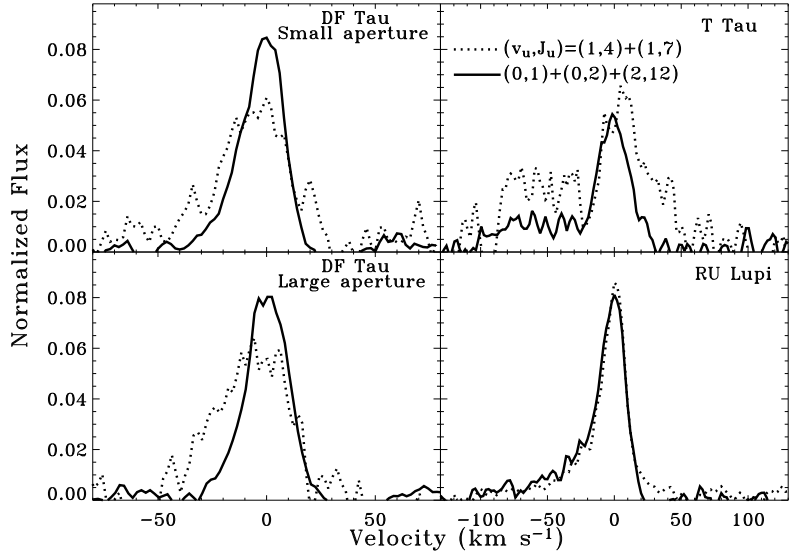}
\caption{Spectral profiles of H$_2$ lines from upper levels (\Vup,\Jup)=(0,1), (0,2), 
and (2,12) (solid lines), which are pumped on the red
wing of \Lya, compared with spectral profiles of H$_2$ emission from
(1,4) and (1,7) (dotted lines), which are pumped near
line center of \Lya.  The lines pumped near \Lya\ line center, from (1,4) and (1,7), 
in the T Tau and DF Tau spectra show excess blueshifted emission, relative
to the flux in the narrow component, than is seen in lines 
from (0,1), (0,2), and (0,12).  The same lines from T Tau also show some excess redshifted emission.
We do not detect any significant difference
in emission line profiles for the different progressions in any other star,
as illustrated here by RU Lupi.} 
\end{figure}

\begin{figure}
\plotone{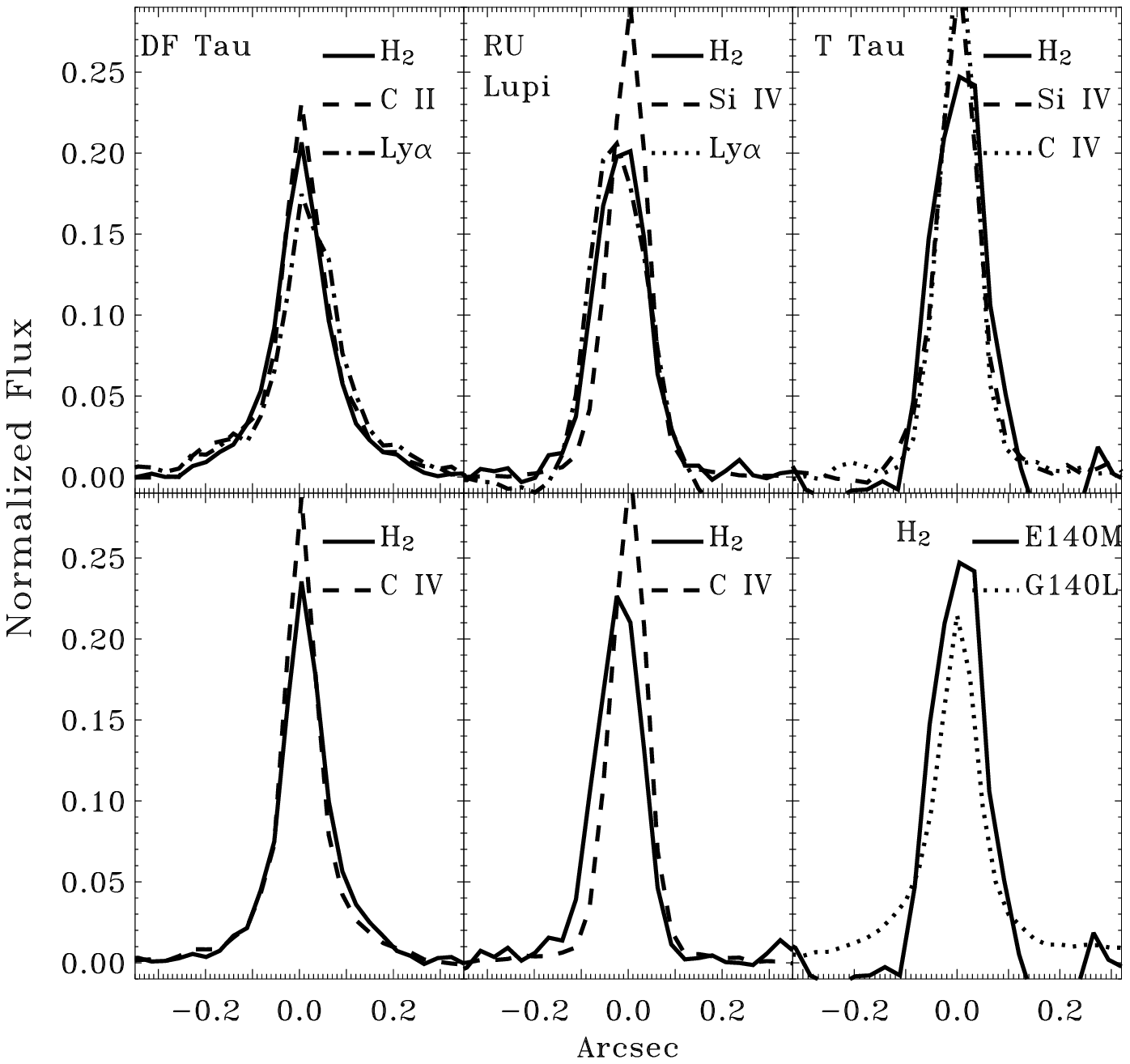}
\caption{The spatial extent of selected emission lines in the cross-dispersion direction.
The top left and center panels compare the spatial profiles of coadded H$_2$ lines between
$1270<\lambda<1400$ \AA\ with \ion{H}{1} \Lya\ 1215.67 \AA, the \ion{Si}{4}
1400 \AA\ doublet, and the \ion{C}{2} 1335 \AA\ doublet, while the bottom left and center
panels compare the spatial profiles of coadded H$_2$ lines between $1500<\lambda<1650$
\AA\ with the \ion{C}{4} 1549 \AA\ doublet.  
For T Tau (right panels), we coadd all H$_2$ lines for comparison with the other emission 
lines to increase S/N and because the point-spread function during that
observation appears constant with wavelength. 
The H$_2$ emission
from DF Tau is not significantly extended, while the H$_2$ and \Lya\
emission are clearly extended in one direction toward RU Lupi.  The H$_2$
emission from T Tau may be slightly extended.}
\end{figure}

\begin{figure}
\epsscale{0.9}
\plotone{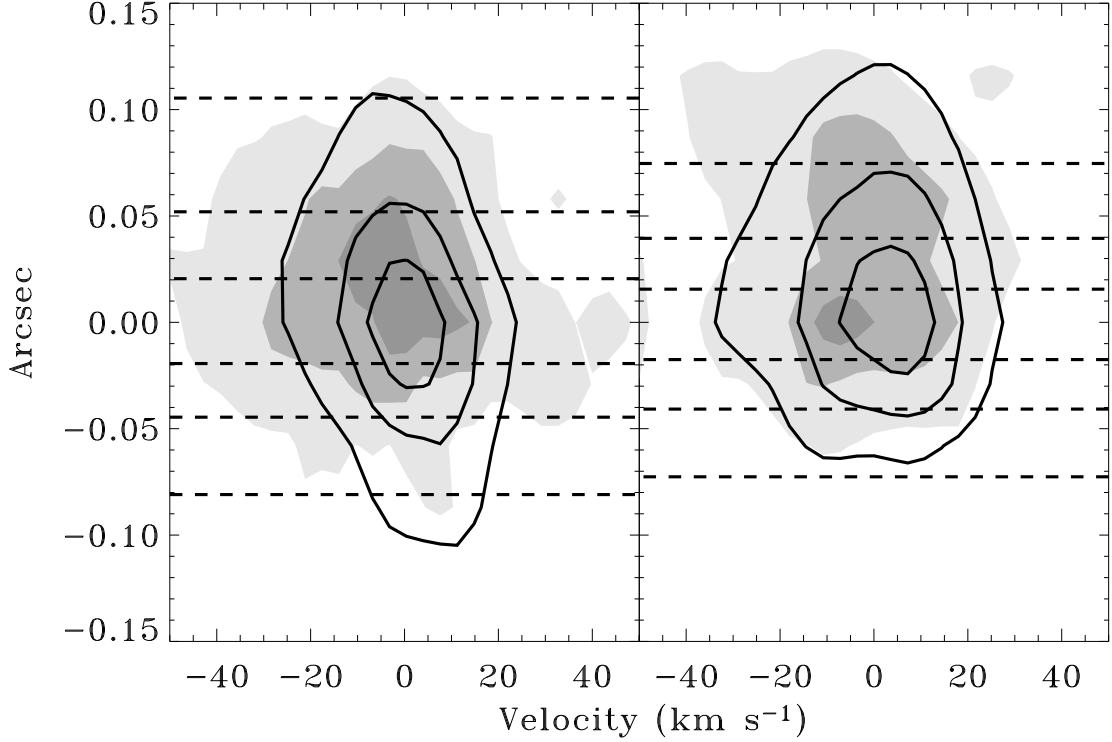}
\caption{The space-velocity diagram of background-subtracted H$_2$ emission from DF Tau for the
large (left) and small (right) aperture observations.  The solid lines are
contours of 0.2, 0.5, and 0.8 times the peak flux for H$_2$ lines from 
(\Vup,\Jup)=(0,1), (0,2), and (2,12), while the shaded regions
are similar contours for  (1,4) and (1,7).  The background was calculated by coadding 
background regions nearby each of the H$_2$ lines.  The contours of \ion{C}{2} emission 
(dashed lines on left) and \ion{C}{4} emission (dashed lines on right) indicate the 
point-spread function during the two observations.
The H$_2$ lines from the progressions pumped near line center [(1,4) and (1,7)] are clearly
blueshifted, and also appear to be slightly extended to the SW (up in this figure).}
\end{figure}

\begin{figure}
\epsscale{0.9}
\plotone{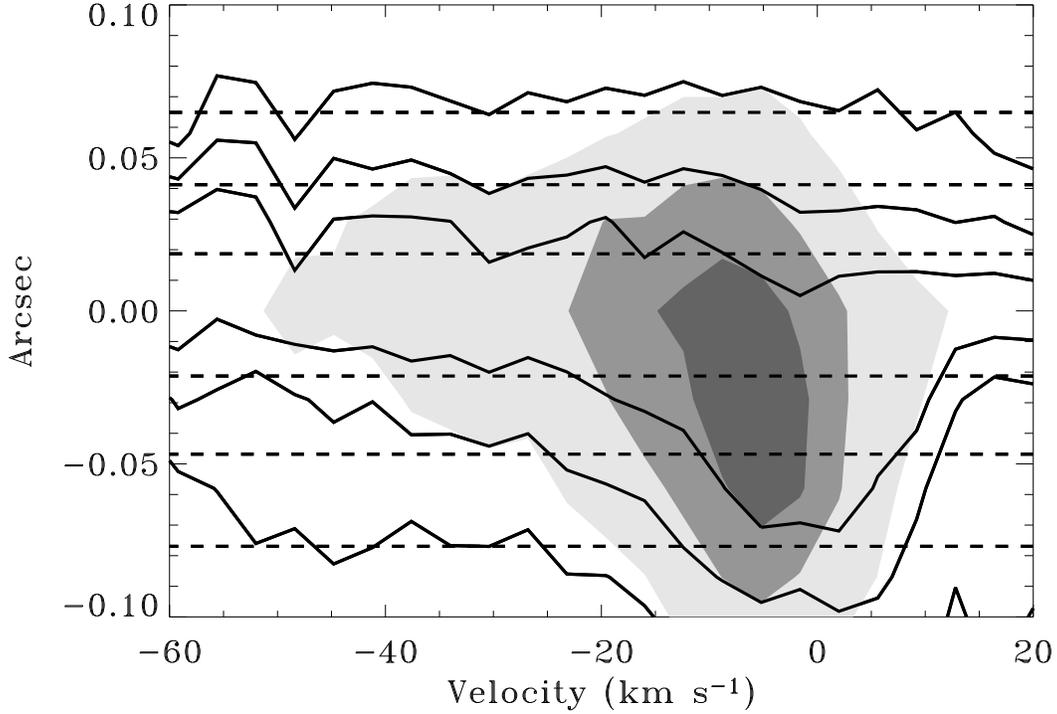}
\caption{The space-velocity diagram of background-subtracted H$_2$ emission from RU Lupi.  
Flux contours of  0.2, 0.5, and 0.8 times the peak flux in any
pixel (shaded regions) show where most of the H$_2$ emission occurs.  The flux contours 
of 0.2, 0.5, and 0.8 times the peak flux in a pixel at that
particular velocity (solid lines) display the spatial extent of emission across the line 
profile.  The dashed lines represent the point-spread function measured from \ion{Si}{4} 
emission.  These contours show that the H$_2$
emission at $v=-15$ \kms\ is spatially extended to the SW (down in this
figure).  However, the H$_2$ emission at $v<-25$ \kms\ is not extended
beyond a point source.  The spatial profile at $v=-15$ \kms\ is consistent
with 70\% of the H$_2$ originating on-source and 30\% of the emission
either being produced smoothly across the aperture or in a point source, located at
$\sim70$ mas from the star.}
\end{figure}

\begin{figure}
\plotone{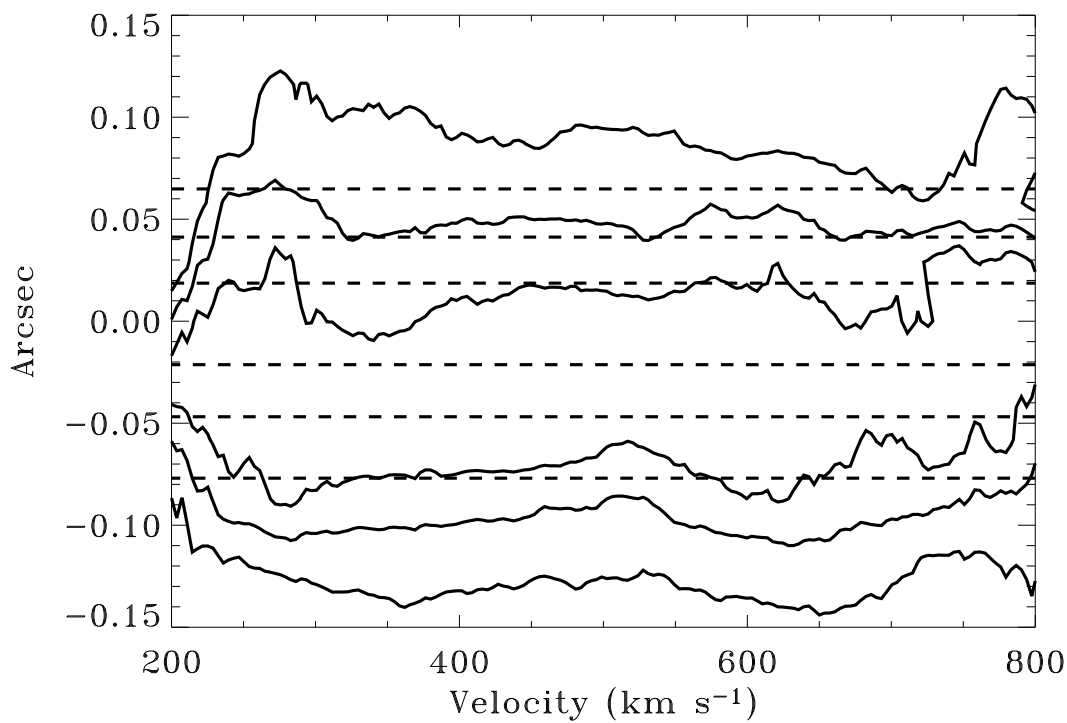}
\caption{The spatial extent of \Lya\ emission toward RU Lupi, smoothed by
75 \kms.  The solid lines show the flux contours of 0.2, 0.5, and 0.8 times the peak
emission at that velocity.  The dashed lines show the point-spread function
of the observation, measured from \ion{Si}{4} emission.  The \Lya\ emission
is clearly extended in the SW direction (down in this figure), which matches the spatial extent
of the H$_2$ emission.  The amount of spatially extended emission does not
depend on velocity from line center.}
\end{figure}

\begin{figure}
\plotone{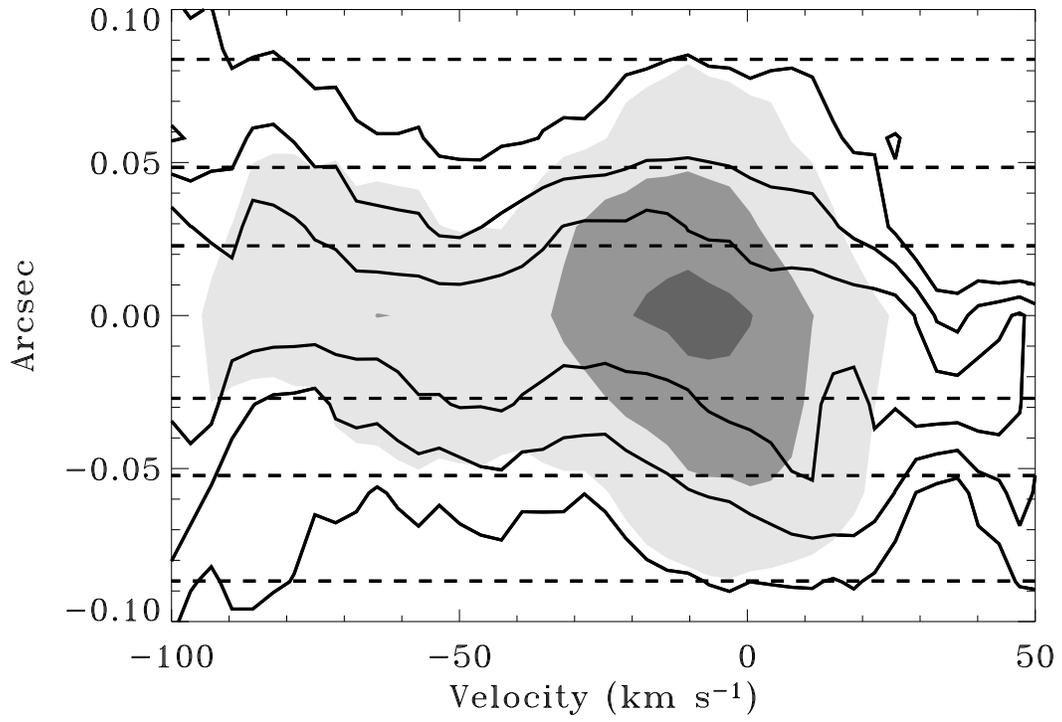}
\caption{Same as Fig. 14, except for T Tau.  
The edge of the plot, particularly at $v=40$ \kms, is dominated by noise and unreliable.}
\end{figure}

\begin{figure}
\epsscale{0.6}
\plotone{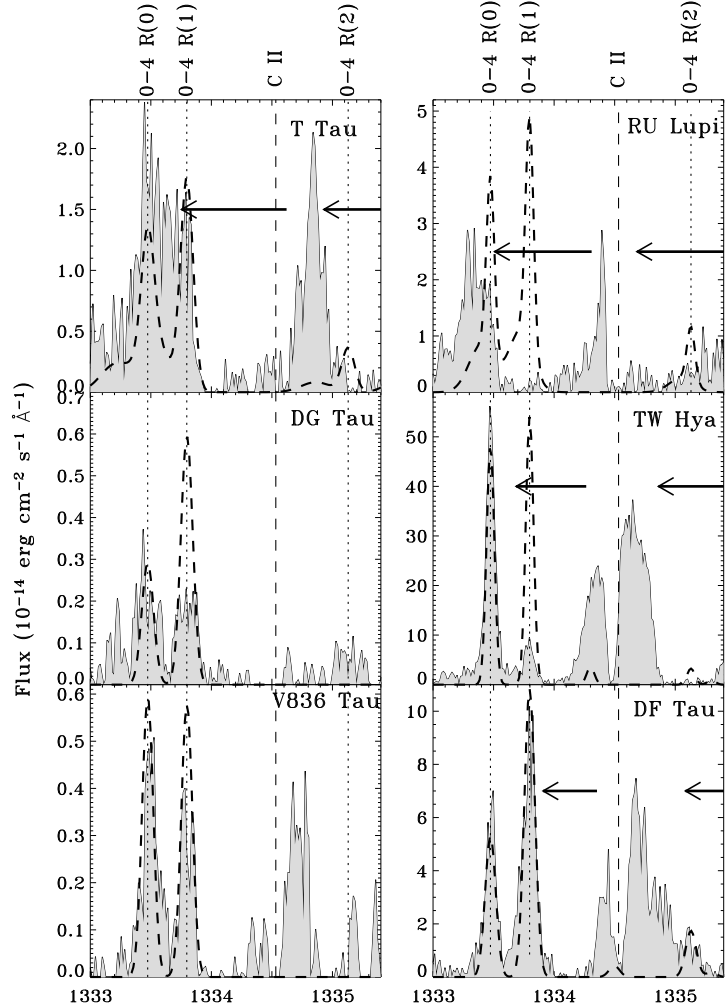}
\caption{Same as Figure 4, except expanding a narrow spectral region.
\ion{C}{2} wind absorption from some sources, 
such as T Tau and RU Lupi, attenuates much 
of the otherwise strong \ion{C}{2} emission and several H$_2$ lines.
 The 0-4 R(0), 0-4 R(1), and 0-4 R(2) lines from
RU Lupi are not present, and the 0-4 R(1) and 0-4 R(2) lines from TW Hya are
much weaker than expected because of wind absorption.  The wind of DF Tau is not optically thick in \ion{C}{2} at $<-140$ \kms,
so both of these H$_2$ lines are as strong as
predicted from our models (see \S 4.3).  In the T Tau 
spectrum, the 0-4 R(1) line is weaker than predicted,
assuming some \ion{C}{2} emission at that wavelength, and
the 0-4 R(2) line may also not be present because of wind
absorption.  The wind our line of sight to V836 Tau is not
expected to be optically thick at the velocities of these H$_2$ lines.  The 0-4
R(1) line to DG Tau appears weaker than expected, but several lines in the
DG Tau spectrum (0-5 R(0) and 0-5 R(1) at 1394 \AA, and 0-6 R(0) and 0-6
R(1) at 1455 \AA) are unexpectedly weak for an unknown reason.  As a
result, we consider any attenuation of the line by the wind of DG Tau
inconclusive.}
\end{figure}

\begin{figure}
\epsscale{0.9}
\plotone{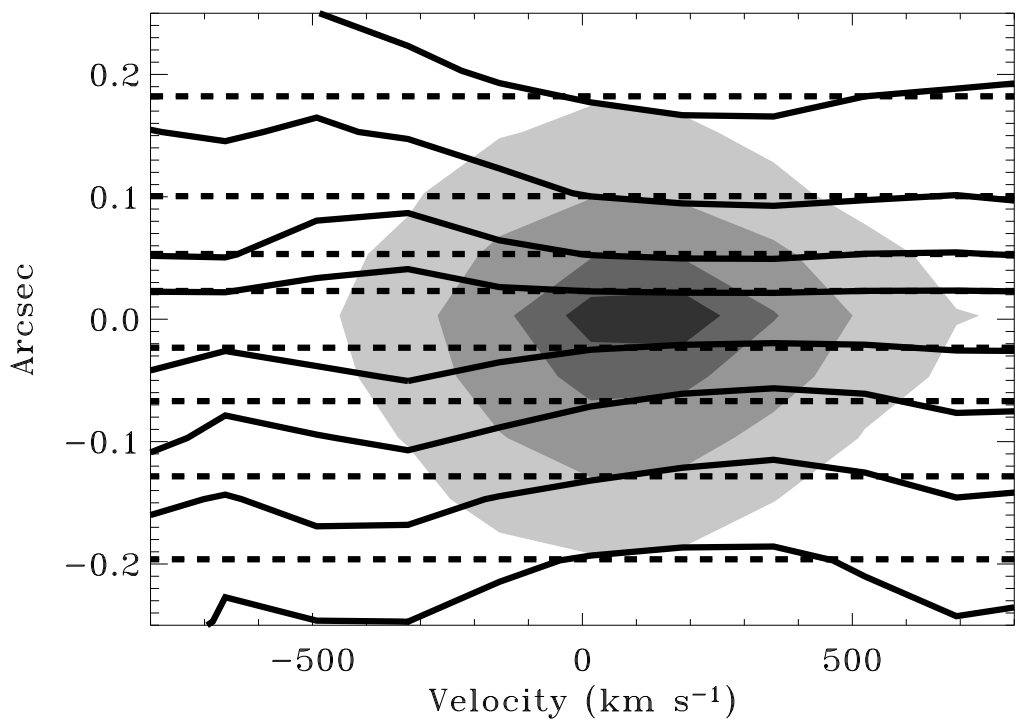}
\caption{Same as Fig. 15 for H$\beta$ in the low-resolution G430L spectrum of RU Lupi, with 
contours of 0.05, 0.2, 0.5, and 0.8.  The point-spread function (dashed lines) is estimated 
from regions near H$\beta$ that are dominated by the continuum.  The bulk of the H$\alpha$ 
emission may be produced by accreting gas, and appears redshifted presumably because of  wind absorption, which is not detectable at our spectral resolution.  The 
blueshifted H$\beta$ emission may be produced by the stellar wind and is 
extended symmetrically about the star.}
\end{figure}

\clearpage

\begin{deluxetable}{lccccccc}
\tabletypesize{\small}
\tablecaption{LOG OF STIS OBSERVATIONS}
\tablehead{Star & Date  & Exp. Time(s) & Grating & Aperture & Bandpass (\AA)& Spectral
Res. & PA$^a$}
\startdata
DF Tau$^b$ & 9/18/99  &  72 & G430L & $52\farcs\times0\farcs2$ &  2850--5750 & 
1500 & 215\\
DF Tau$^b$ & 9/18/99 & 1670 & E230M & $0\farcs2\times0\farcs2$
& 2130--2810 & 30000 & 215\\
DF Tau$^b$ & 9/18/99  & 2320 & E140M & $0\farcs5\times0\farcs5$ &
1170--1710 & 25000 & 215\\
DF Tau$^c$ & 1/28/00 & 36 & G430L & $52\farcs\times0\farcs1$ & 2850--5750 & 1500
& 38\\
DF Tau$^c$ & 1/28/00 & 9390  & E140M & $0\farcs2\times0\farcs06$ &
1170-1710 & 45000 & 38\\
TW Hya$^d$ & 5/7/00  &  72 & G430L & $52\farcs\times0\farcs2$ &  2850--5750 & 
1500 & 43\\
TW Hya$^d$ & 5/7/00 & 1670 & E230M & $0\farcs2\times0\farcs2$
& 2130--2810 & 30000 & 43\\
TW Hya$^d$ & 5/7/00  & 2320 & E140M & $0\farcs5\times0\farcs5$ &
1170--1710 & 25000 & 43\\
\enddata
\end{deluxetable}

\begin{deluxetable}{lcccccccccc}
\tabletypesize{\footnotesize}
\tablecaption{Stellar Properties}
\rotate
\tablehead{
Star & Sp.& Class & $A_V$ & $L_{acc}$ & $\dot{M}_{acc}$ & $v_r$ & $v \sin i$ & Incl. & Mult.\\ 
 & Type & & (mag) & ($L_\odot$) & ($10^{-8}$ \mdotyr) & (\kms) & (\kms) & ($^\circ$) & (\#)}
\startdata
T Tau N   & K0 & CTTS & 0.3$^a,z$   & $0.9^n$&    $4^{n}$        & $18^k$ & $20\pm5^k$   & 15$^b$& 3$^b$\\
DF Tau   & M0 & CTTS & 0.5$^k,z$   & 0.24$^k$ &   $3,0.2^k$   &15$^k$ & $\sim 15^k$  & 85$^g$ & 2\\
RU Lupi  & K7 & CTTS & 0.07$^j,z$  & 0.35$^j$ & $3^j$ &$-2^l$ &$9.0\pm0.9^l$ & 23$^m$ & 1--2$^m$\\
DG Tau   & K7 & CTTS &1.6$^c$    & 0.22$^k$ &  3$^k$    &17$^g$ &  37$^g$   & 90$^g$ & 3\\
V410 Tau & K4 & WTTS & 0.7$^d$   &N/A   & N/A     &$33^k$ & $\sim 40^k$  &  -& 3\\
V836 Tau & K7 & WTTS & 0.6$^d$   &-     & 0.1$^k$       &$17^k$ & $\sim 15^k$  & - & 1$^f$\\
V819 Tau & K7 & WTTS & 1.5$^{d,e}$ &N/A   & N/A       &$15^k$ & $<10^k$      & -& 1$^f$\\
TW Hya   & K7 & CTTS & 0.0$^i,z$  & 0.034$^i$ & $0.2^{i}$ & 13$^p$&     $<6^g$       & $7\pm1^h$& 1 \\
\hline
\hline
\multicolumn{8}{l}{$^z$Calculated from $N$(H), assuming the interstellar
  relationship}\\
\multicolumn{8}{l}{of \citet{Boh78}, with $R_V=3.1$ \citep {Car89}. } \\
\multicolumn{4}{l}{$^a$\citet{Wal03}} &
\multicolumn{4}{l}{$^b$several references, see text}\\
\multicolumn{4}{l}{$^c$\citet{Gul00}}&
\multicolumn{4}{l}{$^d$\citet{Ken95}}\\
\multicolumn{4}{l}{$^e$\citet{Whi01}} &
\multicolumn{4}{l}{$^f$\citet{Sim95}}\\
\multicolumn{4}{l}{$^g$\citet{Joh01}} &
\multicolumn{4}{l}{$^h$\citet{Qi04}} \\
\multicolumn{4}{l}{$^i$\citet{Her04}} &
\multicolumn{4}{l}{$^j$\citet{Her05}}\\
\multicolumn{4}{l}{$^k$ This work}&
\multicolumn{4}{l}{$^l$\citet{Ste02}}\\
\multicolumn{4}{l}{$^m$\citet{Gah05}}&
\multicolumn{4}{l}{$^n$\citet{Cal04}}\\
\multicolumn{4}{l}{$^p$\citet{Wei00}}\\
\enddata
\end{deluxetable}

\clearpage
\clearpage
\clearpage

\begin{table}
\caption{Properties of H$_2$ Pumping Transitions}
{\small \begin{tabular}{lcccccccccccc}
$\lambda_{pump}$ (\AA) & $v_{trans}$ (\kms) & band & \Vup & \Jup & \Vlo & \Jlo & $f^a$ ($10^{-3}$ ) & \Elo (eV) & $P_{Dis}^b$\\
\hline
\hline
1031.862$^c$ & -14 & C--X & 1 & 3 & 1 & 3 & 28.1 & 0.60 & 0.0 \\ 
1212.426 & -801 &B--X & 1 & 10 & 1 & 11 & 13.3 & 1.36 & 0.0\\
1212.543 & -772 & B--X & 1 & 13 & 1 & 12 & 10.9 & 1.49 & 0.0\\
1213.356 & -571 & B--X & 3 & 13 & 1 & 14 & 20.6 & 1.79 & 0.015\\
1213.677 & -491 & B--X & 4 & 13 & 2 & 12 & 9.33 & 1.93 & 0.050\\
1214.421 & -308 & B--X & 7 & 4  & 4 &  5 & 5.17 & 2.07 & 0.189\\
1214.465 & -297 & B--X & 3 & 16 & 1 & 15 & 23.6 & 1.94 & 0.032\\
\end{tabular}
}\end{table}

\begin{deluxetable}{ccccccccccccccccccc}
\tablecaption{Fluxes in H$_2$ Progressions$^a$}
\tabletypesize{\small}
\rotate
\tablehead{
\colhead{\Vup} & \colhead{\Jup} & \colhead{$\lambda_{pump}$} & $v_{trans}$ & &\multicolumn{2}{c}{DF Tau} & \multicolumn{2}{c}{DF Tau} & \multicolumn{2}{c}{RU Lupi} & \multicolumn{2}{c}{T Tau} & \multicolumn{2}{c}{DG Tau} & \multicolumn{2}{c}{V836 Tau}& \multicolumn{2}{c}{TW Hya}\\
& &(\AA) &(\kms) && \# & $F_{obs}$ &  \# & $F_{obs}$ & \# & $F_{obs}$ & \# & $F_{obs}$ & \# & $F_{obs}$ & \# & $F_{obs}$ & \# & $F_{obs}$}
\startdata
1 & 10 & 1212.426 & -801 & & - &  -   &  -& -    & - & -  & - &  - & - & - & - & - & 9 & 4.2\\
1 & 13 & 1212.543 & -772 & & - &  -   &  -& -    & - & -    & - &  - & - & - & - & - & 5 & 1.6\\
3 & 13 & 1213.356 & -571 & &  - &  -   &  -& -    & - & -   & - &  - & - & - & - & - & 8 & 4.7\\
4 & 13 &1213.677 & -491 & & - &  -   &  -&  -    & - & -    & - &  - & 4 & 0.10 & - & - & 8 & 2.4\\
7 & 4  & 1214.421 &-308 & &  - &  -   &  -& -    & - & -   & - &  - & - & - & - & - & 3 & 0.85\\
3 & 16 &1214.465 & -297 && 4  & 0.52 & - & -     & 7 & 0.97& - & - & 6 & 0.19 & - & - & 13 & 14.9\\
4 & 4  &1214.781 & -219 & & 5 & 0.76 & 2 & 0.25  & 9 & 1.8& - & - & 3 & 0.099 & - & - & 12 & 8.9\\
4 & 7  &1214.995 &-167 & & - &  -   &  -&  -    & 6 & 0.82  & - & -  & - &  -   & 2 & 0.21 & 9 & 2.6\\
\enddata
\end{deluxetable}

\begin{deluxetable}{cccccccccc}
\tabletypesize{\footnotesize}
\tablecaption{LYMAN-BAND H$_2$ LINES$^a$, LISTED BY PROGRESSION$^b$}
\tablehead{
\colhead{ID} & \colhead{$\lambda_{calc}$ (\AA)} & \colhead{$f_{br}^c$} & \colhead{TW} & \colhead{DF$^d$} & \colhead{DF$^e$} & \colhead{RU} & \colhead{T} & \colhead{DG} & \colhead{V836}\\
\colhead{} &\colhead{} &\colhead{} &   \colhead{Hya} & \colhead{Tau} & \colhead{Tau} & \colhead{Lupi} & \colhead{Tau} & \colhead{Tau} & \colhead{Tau}}
\startdata
          \multicolumn{10}{c}{\Vup=1, \Jup=10, pumped by 1-1 P(11) at 1212.426 \AA}\\
\hline
      1-2 P(11) &     1267.222 & 0.084 &    2.2 &    - &    - &    - &    - &    - &    - \\
       1-3 R(9) &     1292.448 & 0.088 &    4.9 &    - &    - &    - &    - &    - &    - \\
      1-3 P(11) &     1323.232 & 0.076 &    3.8 &    - &    - &    - &    - &    - &    - \\
       1-4 R(9) &     1348.619 & 0.034 &    2.3 &    - &    - &    - &    - &    - &    - \\
       1-6 R(9) &     1462.144 & 0.051 &    3.8 &    - &    - &    - &    - &    - &    - \\
      1-6 P(11) &     1494.193 & 0.080 &    4.8 &    - &    - &    - &    - &    - &    - \\
       1-7 R(9) &     1518.134 & 0.106 &    5.6 &    - &    - &    - &    - &    - &    - \\
\enddata
\end{deluxetable}

\begin{table}
\tabletypesize{\footnotesize}
\caption{HIGHLY EXCITED H$_2$ LINES$^a$, LISTED BY PROGRESSION$^b$}
\begin{tabular}{ccccc}
\hline
ID & $\lambda_{calc}$ (\AA) & $f_{br}^c$ & TW Hya & DF Tau\\
\hline
\multicolumn{5}{c}{\Vup=0, \Jup=24, pumped by 0-3 P(25) at 1547.971 \AA}\\
\hline
      0-2 P(25) &     1498.007 & 0.049 &    2.9 &    1.0 \\
      0-4 R(23) &     1542.241 & 0.168 &    4.0 &    1.3 \\
      0-4 P(25) &     1594.059 & 0.180 &    5.4 &    2.1 \\
      0-5 R(23) &     1586.688 & 0.168 &    - &    1.5   \\
\hline 
   \multicolumn{3}{c}{Total Observed Flux} &    12.3 &     5.9 \\
  \multicolumn{3}{c}{Extinction Corrected Flux} &    12.3 &    19.6\\
\end{tabular}
\end{table}

\clearpage
\clearpage

\begin{deluxetable}{ccccccccccc}
\tabletypesize{\footnotesize}
\tablecaption{IDENTIFICATION AND FLUXES$^a$ OF BLENDED H$_2$ LINES}
\tablehead{
$\lambda$ & ID1 & ID2 &  TW & DF$^b$ & DF$^c$ & RU & T & DG & V836\\
  & & & Hya & Tau & Tau & Lupi & Tau & Tau & Tau}
\startdata
1266.9 & 4-4 P(5)  & 4-1 P(19)   & 13.1 & 1.3 & 0.40 & -  & - & - & -\\
1320.2 & 3-2 P(18) & 5-7 P(8)    & -    & 2.2 &  -   & -  & - & - & -\\
1408.8 & 1-5 P(8)  & 3-4 P(17)   &  -   &  -  &  -   & 1.8& - & - & -\\    
1429.7 & 2-5 R(14) & 3-5 R(16)   & 4.6  & 1.5 &  -   & -  & - & - & -\\
1454.9 & 0-6 R(0)  & 0-6 R(1)    &  -   &  -  &  -   & -  & 4.0 & 0.51 & -\\
1493.7 & 0-4 P(18) & 0-3 R(23)   & 4.7  & 0.8 &  -   & -  & - & - & -\\
1516.2 & 0-7 R(0)  & 0-7 R(1)    & 22.0 & 5.4 & 1.7  & 3.7& 1.3 & 0.59 & -\\
1522.8 & 3-7 P(14) & 3-7 R(16)   & 5.2  &  -  &  -   & -  &- & - & -\\
1572.7 & 4-10 P(5) & 2-8 P(11)   & 8.3  &  -  &  -   & -  &- & - & -\\
1576.9 & 2-8 R(14) & 0-8 R(0)$^d$& 15.1 &  -  &  -   & -  &- & - & -\\
1585.8 & 4-10 P(8) & 0-8 P(3)    & 7.4  & 3.9  & -   & -  &- & - & -\\
1586.8 & 4-10 P(12)& 0-5 P(23)   & 10.8 &  -  &  -   & -  &- & - & -\\
\hline
\multicolumn{5}{l}{$^a$Flux in $10^{-15}$ erg cm$^{-2}$ s$^{-1}$}\\
\multicolumn{5}{l}{$^b$Data obtained with the $0\farcs5\times0\farcs5$ aperture}\\
\multicolumn{5}{l}{$^c$Data obtained with the $0\farcs2\times0\farcs06$ aperture}\\
\multicolumn{5}{l}{$^d$Also 0-8 R(1)}\\
\enddata
\end{deluxetable}

\begin{deluxetable}{cccccl}
\tabletypesize{\footnotesize}
\tablecaption{IDENTIFICATION AND FLUXES$^a$ OF WERNER-BAND H$_2$ LINES}
\tablehead{$\lambda$ & ID &  TW & DF$^b$ & RU & Pump\\
  & & Hya & Tau & Lupi &}
\startdata
1154.910   & 1-3 P(11) & 12.3   & -    &  -  & 1-5 R(9)  1216.997 \AA \\
1172.0$^c$ & 1-4 R(5)  &  7.5   & 2.9 &   - & 1-5 P(5)  1216.988 \AA \\
1174.4$^d$ & 1-3 R(9)  &    -    & 5.6 &  -  & 1-5 R(9)  1216.997 \AA \\
1186.226   & 1-4 R(12) & 4.9    &  -   &  -  & 1-4 P(14) 1214.566 \AA \\
1208.932   & 1-5 Q(3)  &   -     &  -   &2.2 & 1-1 Q(3)  1031.862 \AA\\
1228.406   & 2-6 P(5)  &   -     & 1.3 &  -  & 2-6 P(3)  1217.488 \AA \\
1255.507   & 1-5 P(14) &   5.x8  &  -   & -   & 1-4 P(14) 1214.566 \AA\\
\hline
\multicolumn{6}{l}{$^a$Flux in $10^{-15}$ erg cm$^{-2}$ s$^{-1}$.}\\
\multicolumn{6}{l}{$^b$Data obtained with the $0\farcs5\times0\farcs5$ aperture}\\
\multicolumn{6}{l}{$^c$Also 0-3 Q(10), pumped by 0-4 Q(1) 1217.263 \AA}\\
\multicolumn{6}{l}{$^d$Also 2-5 R(1), pumped by 2-6 R(1) 1217.298 \AA, and}\\
\multicolumn{6}{l}{1-4 Q(7), pumped by 1-5 Q(7) 1218.508 \AA}\\
\enddata
\end{deluxetable}

\begin{deluxetable}{ccccccccccc}
\tablecaption{Fluxes in Strong Lines and H$_2$ Progressions$^a$}
\tabletypesize{\footnotesize}
\rotate
\tablehead{
\colhead{Star} & \colhead{\ion{C}{4}} & \colhead{\ion{Si}{4}} &
\colhead{\ion{C}{2}} & \colhead{\ion{O}{1}} & \colhead{\ion{He}{2}} &
\colhead{H$_2$} & \colhead{H$_2$} & \colhead{H$_2$} & \colhead{H$_2$} & \colhead{H$_2$}\\
\colhead{} & \colhead{1549 \AA} & \colhead{1400 \AA} & \colhead{1335 \AA} & \colhead{1305 \AA} &\colhead{1640 \AA} & \colhead{(1,4)} & \colhead{(1,7)} & \colhead{(0,1)} & \colhead{(0,2)} & \colhead{(2,12)}}
\startdata
TW Hya   & 284& 11 & 11 & 95 & 1.4 & 38 & 17 &
41 & 43 & 19\\
RU Lupi  & 70 & 69 & 11 & 26 & 7.6 & 9.2 & 4.7 &
5.4 & 6.1 & 3.1\\
DF Tau (large)  & 33 &  2.5 & 4.3 & 5.6 & 8.1 & 5.2 & 
2.2 & 7.0 & 12.3 & 8.4\\
DF Tau (small)  & 2.9 &  0.38 & 0.78 & 1.6& 0.82 & 1.6 & 
0.89 & 2.4 & 4.2 & 3.0\\
T Tau    & 22.6 & 5.6 & 2.5 & 25 & 5.5 & 1.8 & 1.6 &
2.5 & 3.8 & 2.0\\
DG Tau   &$<0.08$ & $<0.05$& $<0.04$ & 0.09 & $<0.07$ & 0.30 & 0.19 &
0.48 & 0.86 & 0.46\\
V836 Tau & 1.4 & 0.13 & 0.38 & 0.44 & 0.84 & 3.9 & 1.6 & 4.2 & 
3.5 & 2.1\\
V410 Tau & 2.8 & 0.88& 1.2 & 1.1 & 1.7 & $<0.3$ & $<0.16$
& $<0.08$ & $<0.2$ & $<0.2$\\
V819 Tau & $<0.2$ & $<0.2$ & $<0.1$ & $<0.2$ & $<0.08$ & $<0.2$ & $<0.2$ & $<0.07$ & $<0.2$ & $<0.2$ \\
\hline
\multicolumn{8}{l}{$^a10^{-14}$ erg cm$^{-2}$ s$^{-1}$, not corrected for extinction}\\
\enddata
\end{deluxetable}

\begin{deluxetable}{lcccccccc}
\tablecaption{Multiple components of H$_2$ Emission}
\tablehead{
Star & (\Vup, \Jup) & Component$^a$ & $v_{{\rm H}2}$ & FWHM &  Flux Fraction$^e$}
\startdata
DF Tau$^b$ & (0,1);(0,2);(2,12) & 1 & -1 & 23 & N/A\\
DF Tau$^b$ & (1,4);(1,7) & 1 & -8 & 41 & N/A\\
DF Tau$^{c,d}$ & (0,1); (0,2); (2,12) & 1 & 6 & 19 & 0.70\\
       & (0,1); (0,2); (2,12) & 2 & -8 & 30 & 0.30 \\
DF Tau$^{c,d}$ & (0,1); (0,2); (2,12) & 1 & 3 & 24 & N/A\\
DF Tau$^c$ & (1,4);(1,7) & 1 & 0 & 36 & N/A\\
RU Lupi & all  & 1 & -12 & 18 & 0.57\\
        & all  & 2 & -30 & 53 & 0.43 \\
T Tau   & (0,1); (0,2) & 1 & -12 & 28 & 0.64\\
        &  (0,1); (0,2) & 2   & -72 &65 & 0.36\\
 T Tau  & (1,4); (1,7) & 1 & -2 & 30 & 0.49\\
        & (1,4); (1,7)  & 2  & -65 & 69 & 0.51\\
\hline
\multicolumn{8}{l}{$^a$Component of fit, if the fit used multiple Gaussians.}\\
\multicolumn{2}{l}{$^b$Large Aperture} &\multicolumn{4}{l}{$^c$Small Aperture} \\
\multicolumn{8}{l}{$^d$Lines fit as both a single and double Gaussian.}\\
\multicolumn{8}{l}{$^e$Fraction of flux in a component, where applicable}\\
\enddata
\end{deluxetable}

\begin{deluxetable}{lcccccccc}
\tabletypesize{\footnotesize}
\tablecaption{Global Properties of H$_2$ Emission}
\tablehead{
Star & $v_{H2}^a$ (\kms) & FWHM$^a$ (\kms)& \# Upper Levels & \# Lines$^b$ & $F_{obs}^c$ & $F_{ext}^c$ & $F_{mod}^c$}
\startdata
DF Tau$^d$& -1 &23.2  & 22 & 164 & 51.0 & 188 & 210\\
DF Tau$^e$& 6 & 18.6 & 13 & 94 & 14.4 & 53.5 & 62.1\\
RU Lup   & -12   &  17.6 & 13 & 88 & 28.5 & 34.3 & 48.3\\
T Tau    & -12  &   28.4 &  7 & 41 & 9.26 & 20.2 & 29.2\\
DG Tau   & -27  & 26.1 & 10 & 58 & 2.55 & 152$^f$ & 210\\
V836 Tau &  0 & 24.3 & 6 & 45 & 3.13 & 14.8 & 17.7\\
TW Hya   &  0 & 18.2 & 24 & 209 & 217& 217 & 244\\
\hline
\multicolumn{8}{l}{$^a$Velocity with respect to the star and FWHM are for narrow component only.}\\
\multicolumn{8}{l}{$^b$Several blends from each star are counted as a single line.}\\
\multicolumn{8}{l}{$^c$Observed flux, flux corrected for extinction, and
model flux in $10^{-14}$ erg cm$^{-2}$ s$^{-1}$.}\\
\multicolumn{4}{l}{$^d$Large apeture} & \multicolumn{4}{l}{$^e$Small aperture}\\
\multicolumn{8}{l}{$^f$Extinction to H$_2$ emission from DG Tau is very uncertain.}\\
\enddata
\end{deluxetable}
\begin{deluxetable}{lccccccc}
\tablewidth{0pt}
\tabletypesize{\small}
\rotate
\tablecaption{Attenuation of H$_2$ lines by the wind}
\tablehead{
\colhead{Star} & \colhead{$v_\infty^a$} (\ion{C}{2} 1334.5 \AA)  & \colhead{$v_\infty^a$} (\ion{Mg}{2} 2796 \AA)&
\colhead{$v_{H2}$} & 
\multicolumn{2}{c}{0-4 R(0), -240 \kms} & \multicolumn{2}{c}{0-4 R(1), -165 \kms} \\
 & \colhead{\kms} & \colhead{\kms} &\colhead{\kms} &
Model$^b$ & Obs.$^b$ &  Model$^b$ & Obs.$^b$}
\startdata
DF Tau$^c$ & -190  & -130 & -1 & 6.0 &6.8 & 12.2 & 12.3\\
TW Hya & -205  & 180 & 0 & 36.9 & 42.8 & 42.7 & 7.9\\
RU Lupi & -350$^d$ & -265 & -12 & 4.7 & $<2.1$ & 6.1& $<0.2$ \\
T Tau & -300$^d$  & -180 & -12 & 2.0 &$<3.0$& 3.6& $<2.7$\\
DG Tau & -330$^d$ & - & -27 &0.4 &0.5& 0.8 & $\sim0.4^f$\\
V836 Tau & - & $>-190^e$ & 0 &0.7 &0.8& 0.7 & 0.5\\
\hline
\multicolumn{7}{l}{$^a$Maximum velocity of wind absorption}\\
\multicolumn{7}{l}{$^b$Flux, $10^{-15}$ erg cm$^{-2}$ s$^{-1}$}\\
\multicolumn{7}{l}{$^c$Obtained with large aperture}\\
\multicolumn{7}{l}{$^d$From \citet{Ard02b}}\\
\multicolumn{7}{l}{$^e$Inferred from H$_2$ lines.}\\
  \multicolumn{7}{l}{$^f$Flux estimate unreliable.}\\
  \enddata
\end{deluxetable}
\begin{deluxetable}{lccccc|ccccc}
\tablewidth{0pt}
\tabletypesize{\small}
\rotate
\tablecaption{Comparisons to previous observations}
\tablehead{
& \multicolumn{5}{c}{STIS} & \multicolumn{5}{c}{GHRS$^a$}\\
\colhead{Star} & \colhead{$v_{H2}^b$} & \colhead{FWHM$^b$} & \colhead{$F_{red}^{c,d}$} & \colhead{$F_{cen}^{c,e}$} & \colhead{$F($\ion{C}{4})$^c$} &  \colhead{$v_{H2}^b$} & \colhead{FWHM$^b$} & \colhead{$F_{red}^{c,d}$} & \colhead{$F_{cen}^{c,e}$} & \colhead{$F($\ion{C}{4})$^c$}}
\startdata
\hline
DF Tau$^f$  & -1  & 23 & 46.3$^g$&  5.5$^h$ & 330 & -1   & 27 & 52$^g$ & 4$^h$  & 498\\
RU Lupi & -12 & 18 &  -      & 14.9$^i$ & 700 & -17  & 37 & -      & 26$^i$ & 146\\
T Tau   & -12 & 28 &16.1$^j$ & 12.8$^k$ & 230 & -9   & 52 & 33$^j$ & 98$^k$ & 166\\
DG Tau  & -27 & 26 & 1.9$^l$ & 1.0$^m$ &$<0.8$& -11  & 41 & 19$^l$ & 8$^m$ & 44\\
\hline
\multicolumn{11}{l}{$^a$From \citet{Ard02a}.  DF Tau and RU Lupi were observed with GHRS before COSTAR}\\
\multicolumn{11}{l}{was installed, leading to a $\sim 20$ \kms\ uncertainty in the absolute velocity scale.}\\
\multicolumn{11}{l}{T Tau and DG Tau were observed after COSTAR was installed, and have velocities measured to $\sim 3$ \kms.}\\
\multicolumn{2}{l}{$^b$\kms} & 
\multicolumn{4}{l}{$^c$Flux in various lines, $10^{-15}$ erg cm$^{-2}$ s$^{-1}$}\\
\enddata
\end{deluxetable}
\clearpage
\begin{deluxetable}{lcccccccccc}
\tabletypesize{\footnotesize}
\rotate
\tablecaption{Summary of stellar and H$_2$ emission line properties}
\tablehead{
Star & Class & $A_V^a$ & $i^a$ & $\dot{M}_{acc}^a$ & Origin & $v_{pump}$ range$^b$ &
$v_{H2}^c$ & $v_{blue}^d$ & Extended?\\
& & (mag) & ($^\circ$) & ($10^{-8} \mdotyr$) & (D/W)$^e$ & (\kms) & (\kms)
& (\kms) & Y/N}
\startdata
TW Hya & CTTS & 0.0  & 7 & 0.2 & D & -801 to 913 &  0  & N/A & No\\
DF Tau & CTTS  & 0.5  & 85 & 3,0.2 & D,W & -297 to 1114 & 2.5 & -35 & No\\
RU Lupi & CTTS & 0.07 & 23 & 3 & W & -297 to 913 & -12 & -100 & Yes\\
T Tau & CTTS   & 0.3  & 15 & 4 & W & 14 to 913    & -12 & -100 & No\\
DG Tau & CTTS  & 1.6  & 90 & 3 & W & -491 to 704 & -27 & N/A & Yes?\\
V836 Tau & WTTS$^f$ & 0.6 & ? & 0.1 &D & -167 to 551  & 0   & N/A &
?\\
V410 Tau & WTTS & 0.7 & - & 0& \multicolumn{5}{c}{Possibly significant
  non-detection of H$_2$ emission}\\
V819 Tau & WTTS &1.5 &  - & 0& \multicolumn{5}{c}{Insignificant
  non-detection of H$_2$ emission}\\
\hline
\multicolumn{7}{l}{$^a$See Table 2 for references}\\
\multicolumn{7}{l}{$^b$Range of velocities for pumping transitions, measured from \Lya\ line center.}\\
\multicolumn{7}{l}{$^c$Velocity of narrow component of H$_2$ emission}\\
\multicolumn{7}{l}{$^d$Maximum velocity of asymmetric blueshifted wing on
  H$_2$ emission profiles}\\
\multicolumn{7}{l}{$^e$Disk (D) or Wind (W)}\\
\multicolumn{7}{l}{$^f$Classified as WTTS, but is a weak accretor with a disk.}\\
\hline
\enddata
\end{deluxetable}

\end{document}